\documentclass[aip,cha,reprint]{revtex4-1}

\usepackage[english]{babel}
\usepackage{helvet,graphicx,mathrsfs}
\usepackage{dsfont}

\usepackage{amsmath}
\usepackage{amsfonts}
\usepackage{amssymb}
\usepackage{wasysym}
\usepackage{dsfont}
\usepackage{color,mathrsfs}

\begin{document}

\title{Nonlinear transient waves in coupled phase oscillators with inertia}

\author{David J. J\"org}\email{djoerg@pks.mpg.de}
\affiliation{Max Planck Institute for the Physics of Complex Systems, Dresden, Germany}

\begin{abstract}
\noindent
Like the inertia of a physical body describes its tendency to resist changes of its state of motion, inertia of an oscillator describes its tendency to resist changes of its frequency.
Here we show that finite inertia of individual oscillators enables  nonlinear phase waves in spatially extended 
 coupled systems.
Using a discrete model of coupled phase oscillators with inertia, we investigate these wave phenomena numerically,  complemented by a continuum approximation that permits the analytical description of the key features of wave propagation in the long-wavelength limit.
The ability to exhibit traveling waves is a generic feature of systems with finite inertia and is independent of the details of the coupling function.
\end{abstract}%


\date{\today}

\maketitle

\begin{quotation}
Coupled oscillators can exhibit a variety of dynamical phenomena, from self-organized synchronization to the formation of complex phase patterns. 
In particular, coupling can enable traveling waves in spatially extended oscillator systems.
Whether phase waves are possible in such a system usually depends on the details of the coupling function.
We here show that if the individual oscillators respond slowly to external coupling signals---a property often called `inertia'---, wave phenomena appear as a generic feature of the system, independent of the details of the coupling function.
We investigate these wave phenomena numerically and analytically and show how the properties of the coupling function affect wave propagation in typical scenarios.
\end{quotation}

\section{Introduction}

\noindent%
Coupled oscillators are fascinating entities, not only because of their robust synchronization properties but also as elementary constituents of pattern forming systems \cite{Pikovsky2003,Kuramoto1984,Cross1993}.
Spatially extended systems of coupled oscillators can exhibit a rich variety of patterns such as waves, solitons, spirals, target patterns, splay states, and other stationary or transient phase distributions \cite{Cross1993,Strogatz1993,Alexeeva2000,Jeong2002,Kim2004,Ahnert2008,Lee2011,Ares2012,Jorg2014}.
Among these patterns, traveling waves play a prominent role as they encode spatiotemporal information in a particularly simple way.
Traveling waves in coupled oscillator systems have been reported, e.g., in continuous reaction--diffusion systems \cite{Ortoleva1974,Cross1993} and in discrete systems, e.g., due to specific types of coupling \cite{Sakaguchi1988,Rosenau2005,Lee2011,Murray2011,Hoang2015} or spatial gradients of the intrinsic frequency of the oscillators \cite{Morelli2009,Ares2012}.
Depending on the mechanism that gives rise to wave propagation, waves can be persistent \cite{Rosenau2005,Ahnert2008} (e.g., in the case of solitons) or transient\cite{Szwaj1996,Matias1998,Horikawa2012} (e.g., in the case of damped waves).
While the crucial properties that enable traveling waves have mostly been sought in the particular way that oscillators interact, we here consider the role of slow frequency adaptation of the individual oscillators for wave phenomena in the coupled system.
This slow frequency adaptation, often termed `inertia' \cite{Tanaka1997a,Acebron1998,Choi2013,Olmi2014}, enters the governing equations of a phase oscillator system as a second time derivative of the phase.
Like the inertia of a mass point describes its tendency to resist changes of its velocity,
inertia of a phase oscillator describes its tendency to resist changes of its dynamic frequency \cite{Ermentrout1991}.
These frequency changes may be induced by external coupling signals or a time-dependent intrinsic frequency of the oscillator.
The role of inertia for synchronization has been extensively studied in recent years \cite{Ermentrout1991,Tanaka1997a,Tanaka1997b,Acebron1998,Hong2002,Dolan2005,Choi2013,Ji2013,Gupta2014,Komarov2014,Olmi2014}.
Inertia introduces qualitatively new effects such as first order phase transitions in the synchronization onset \cite{Tanaka1997a}, cluster explosive synchronization \cite{Ji2013}, and hysteretic transitions from incoherence to coherence \cite{Tanaka1997a,Tanaka1997b,Olmi2014}.
We here consider the influence of inertia (i.e., slow frequency adaptation) on the ability of a coupled oscillator system to exhibit phase waves.

In this paper, we show that finite inertia enables nonlinear damped  phase waves in spatially extended oscillator systems.
To this end, we study a nearest-neighbor coupled lattice of identical phase oscillators.
This is complemented by a spatial continuum approximation for the limit of long wavelengths, which elucidates the mechanism of wave propagation and enables to analytically compute the velocity and decay rate of waves as well as the effects of nonlinear coupling on wave propagation.
The results show that in the presence of inertia, the ability of oscillator systems to exhibit waves does not depend on the specific type of coupling but rather is a generic feature of such systems, which enables signal propagation at finite velocities.

\section{Coupled phase oscillators with inertia}

We study a phase oscillator model for a network of coupled identical oscillators with inertia, introduced by Tanaka, Lichtenberg, and Oishi \cite{Tanaka1997a,Tanaka1997b}, based on the zero-inertia model by Kuramoto \cite{Kuramoto1984},
\begin{align}
	\mu \ddot\phi_i + \dot\phi_i = \Omega + \frac{\kappa}{n_i} \sum_{j=1}^N w_{ij} \Gamma(\phi_j-\phi_i) \ ,\label{eq.phase.model}
\end{align}
where $\phi_i$ is the phase of oscillator $i$, $\mu>0$ is the inertia of the oscillators, $\Omega$ is the intrinsic frequency, $\kappa$ is the coupling strength, $\Gamma$ is a $2\pi$-periodic coupling function, $(w_{ij})$ is the adjacency matrix, and $n_i=\sum_j w_{ij}$ is the total weight of links of oscillator $i$.
We here consider the case of nearest-neighbor coupling on a $d$-dimensional hypercubic lattice, in which case $(w_{ij})$ is diagonalizable and $n_i=2d$.
Without loss of generality, we consider coupling functions $\Gamma$ that satisfy $\Gamma(0)=0$ as any finite $\Gamma(0) \equiv \Gamma_0$ can be absorbed by the redefinition $\Omega \to \Omega + \kappa \Gamma_0$ and $\Gamma(\varphi) \to \Gamma(\varphi) - \Gamma_0$.
The inertia $\mu$ determines the time scale on which an oscillator adapts its frequency: for an  uncoupled oscillator starting with initial frequency $\dot\phi(0)=\Omega_0$ and described by $\mu \ddot\phi + \dot\phi = \Omega$, the time-dependent dynamic frequency is given by $\dot\phi(t)=\Omega+(\Omega_0-\Omega)\mathrm{e}^{-t/\mu}$.

The phase waves described in the following are (possibly large) perturbations of the in-phase synchronized solution
\begin{align}
	\phi_i(t) = \Omega t \ , \label{eq.sync}
\end{align}
in which all oscillators evolve with their intrinsic frequency $\Omega$ and have no phase lag relative to each other.
Hence, before turning to the description of wave phenomena, a comment on the stability of the solution Eq.~(\ref{eq.sync}) is in order.
Linear stability analysis yields that this solution is stable if and only if $\kappa \Gamma'(0)>0$, a result coinciding with the one for homogeneous systems without inertia, see Appendix~\ref{app:stability}.
We only consider such cases.
To simplify the description of waves, we consider the phase dynamics in the corotating frame, $\phi_i \to \phi_i + \Omega t$, which amounts to setting $\Omega=0$ in Eq.~(\ref{eq.phase.model}). 

\begin{figure}[t]
\centering
\includegraphics[width=8.5cm]{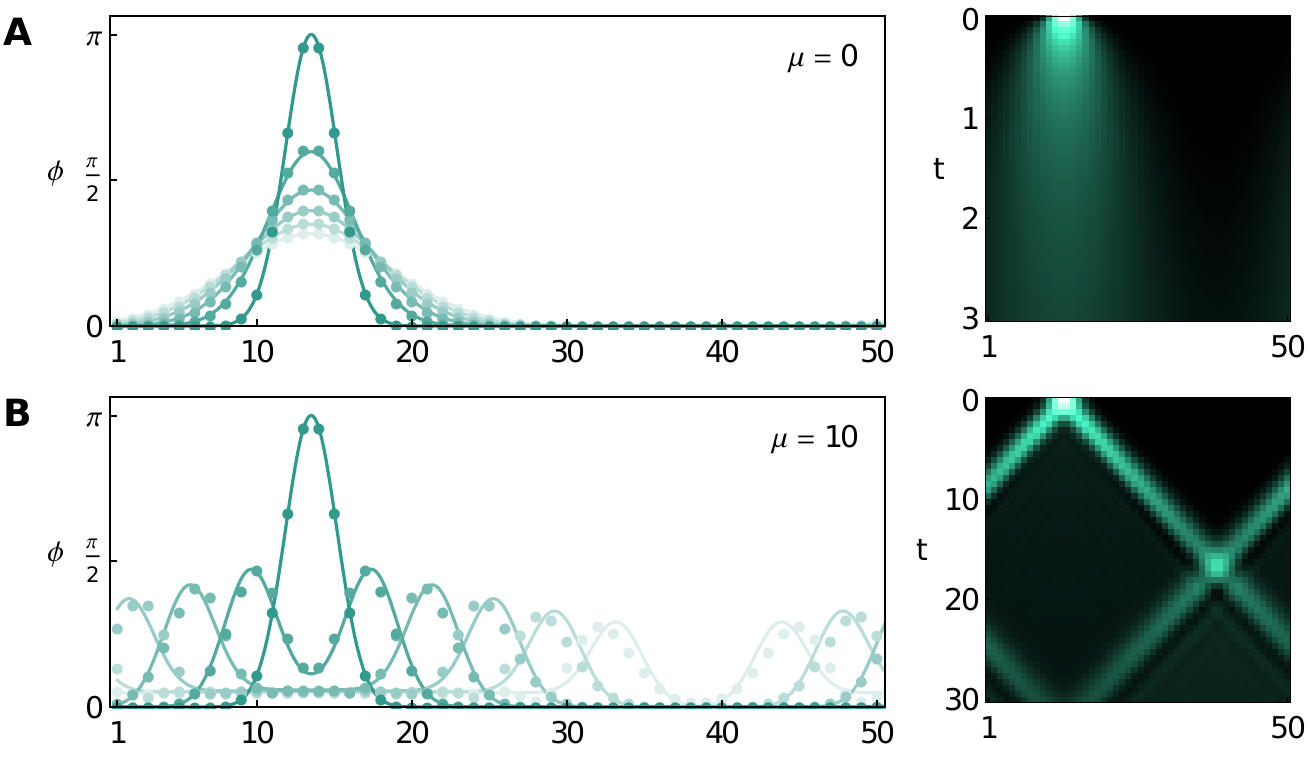}
\caption{Phase distribution of a ring of oscillators for different points in time for the case of (A) no inertia, $\mu=0$, and (B) finite inertia, $\mu=10$. Dots show a numerical solution of the discrete oscillator system Eq.~(\ref{eq.phase.model}); curves show a numerical solution of  the continuum approximation Eq.~(\ref{continuum.1d}). Brighter curves show later times: $t=0,{...},0.5$ in steps of $\Delta t =0.1$ in A; $t=0,{...},12.5$ in steps of $\Delta t =2.5$ in B.
The space-time plots to the right show representations of the same systems in which each pixel along the $x$-axis corresponds to an oscillator. The color value indicates the relative phase of the oscillator (see color wheel in Fig.~\ref{fig2}A) and time proceeds from top to bottom.
The other parameters are $N=50$, $\kappa=50$, and $\Omega=0$. The coupling function is $\Gamma(\varphi)=\sin\varphi$. Initial conditions for Eq.~(\ref{eq.phase.model}): $\phi_i(0)=f(i\varepsilon)$ and $\dot \phi_i(0)=0$ and for Eq.~(\ref{continuum.1d}): $\Phi(x,0)=f(x)$ and $\partial_t\Phi(x,0)=0$, where $f(x)=f_0\exp(-(x-x_0)^2/2 \sigma^2)$ with $f_0=\pi$, $x_0=\ell/4$, and $\sigma=0.2$.}
\label{fig1}
\end{figure}

We now demonstrate that Eq.~(\ref{eq.phase.model}) can exhibit the propagation of damped waves for $\mu\neq 0$.
For simplicity, we here consider a one-dimensional ring of oscillators with nearest-neighbor coupling, $w_{ij}=\delta_{i,j-1}+\delta_{i,j+1}$ with $\delta$ being the Kronecker delta, and sinusoidal coupling function, $\Gamma(\phi)=\sin \phi$.
To show how waves propagate, we start the system with initial phases given by a sharply localized Gaussian distribution in the oscillator index $i$, which approximately satisfies periodic boundary conditions, and solve Eq.~(\ref{eq.phase.model}) numerically.
Fig.~\ref{fig1} shows the evolution of this initial phase distribution in a system with no inertia ($\mu=0$) and a system with finite inertia ($\mu \neq 0$).
For the case of no inertia (Fig.~\ref{fig1}A), the initial phase distribution decays and broadens in a diffusive manner, while staying centered around its initial position.
For the case of finite inertia (Fig.~\ref{fig1}B), the initial phase distribution initiates a decaying wave with finite velocity in both directions of the ring.
In both cases, the phase differences between oscillators eventually decay and in the limit of large times, the in-phase synchronized state Eq.~(\ref{eq.sync}) is reached.

\section{Continuum approximation of the oscillator lattice}

Wave propagation and decay can be understood using a spatial continuum approximation of Eq.~(\ref{eq.phase.model}), which effectively describes long-wavelength solutions.
We replace the nearest-neighbor coupled oscillator lattice in Eq.~(\ref{eq.phase.model}) by a continuous field, $\phi_{i_1,{...},i_d}(t) \to \Phi(\mathbf{x},t)$, where $(i_1,{...},i_d) \in \mathds{Z}^d$ and $\mathbf{x} = (\varepsilon i_1,{...}, \varepsilon i_d ) \in \mathds{R}^d$, where $\varepsilon$ is the lattice spacing.
Expanding the r.h.s.~of Eq.~(\ref{eq.phase.model}) in a power series in $\varepsilon$ to second order at $\varepsilon=0$ yields the nonlinear partial differential equation, see Appendix~\ref{app:continuum},
\begin{align}
	(\partial_t^2 - c^2 \nabla^2) \Phi +2\gamma \partial_t \Phi - \lambda (\nabla\Phi)^2 = \tilde\Omega \ , \label{continuum.corotating}
\end{align}
where $\tilde\Omega=\Omega/\mu$ and
\begin{align}
\begin{split}
	c &= \sqrt{\frac{\varepsilon^2}{2d} \frac{\kappa}{\mu}\Gamma'(0)}  \ , \quad
	\gamma =\frac{1}{2\mu} \Bigg. \ , \quad
	\lambda = \frac{\varepsilon^2}{2d} \frac{\kappa}{\mu}\Gamma''(0)  \ . 
\end{split}\label{continuum.parameters}
\end{align}
Eq.~(\ref{continuum.corotating}) reveals that the Laplace operator $\nabla^2$ arising from the expansion in the lattice spacing combines with the second time derivative $\partial_t^2$ of the inertial term to a d'Alembert wave operator with wave velocity $c$.
Interpreting Eq.~(\ref{continuum.corotating}) as a wave equation, $\gamma$ represents a damping coefficient and $\lambda$ describes the strength of the nonlinear term $(\nabla \Phi)^2$.
For the one-dimensional ring of coupled oscillators with $\Omega=0$ considered in Fig.~\ref{fig1}, Eq.~(\ref{continuum.corotating}) reduces to
\begin{align}
	(\partial_t^2 - c^2 \partial_x^2) \Phi + 2\gamma \partial_t \Phi - \lambda (\partial_x\Phi)^2 = 0 \ , \label{continuum.1d}
\end{align}
with periodic boundary conditions
\begin{align}
	\Phi(0,t) \, \mathrm{mod} \, 2\pi&= \Phi(\ell,t) \,  \mathrm{mod} \, 2\pi  \ , \label{bc1} \\
	\partial_x\Phi(0,t) &= \partial_x\Phi(\ell,t)\ , \label{bc2}
\end{align}
where $\ell$ is the system length.
Without loss of generality, we here set $\ell=2\pi$, which implies a lattice spacing $\varepsilon=2\pi/N$.
We now use Eq.~(\ref{continuum.1d}) to discuss wave phenomena of the oscillator system in the limit of long wavelengths.

\section{Linear approximation}

In Fig.~\ref{fig1}A, we considered a system without inertia and sinusoidal coupling, implying $\mu=0$ and $\lambda=0$.
Multiplying Eq.~(\ref{continuum.1d}) by $\mu$ and subsequently setting $\mu=0$, we find that in this case it simplifies to a diffusion equation
\begin{align}
	\partial_t\Phi=D\partial_x^2 \Phi
\end{align}
with constant $D=c^2/(2\gamma)$, which explains the diffusion-like decay and broadening of the initial phase distribution \cite{Kuramoto1984b}, shown in Fig.~\ref{fig1}A.
For finite inertia ($\mu \neq 0$) and $\lambda=0$, which corresponds to the case shown in Fig.~\ref{fig1}B, Eq.~(\ref{continuum.1d}) reduces to a telegraph equation,
\begin{align}
	\mathcal{L} \Phi \equiv (\partial_t^2 - c^2 \partial_x^2 + 2\gamma \partial_t) \Phi  = 0 \ , \label{telegraph}
\end{align}
which describes the propagation of damped waves.
The telegraph equation is linear and can be solved analytically for arbitrary initial conditions \cite{Zauderer2006}.
We here briefly recapitulate the properties of damped wave solutions by considering the dispersion relation of a plane wave $\Phi(x,t)=\mathrm{e}^{\mathrm{i}(kx-z t)}$, given by $z_k = \sqrt{k^2c^2 -\gamma^2} - \mathrm{i}\gamma$, where $\omega_k \equiv \mathrm{Re}\,z_k$ determines the frequency and $r_k \equiv \mathrm{Im}\,z_k$ the decay rate of the wave. The group velocity of a wave packet with wavenumbers sharply centered around the central wavenumber $k_0>\gamma/c$ is given by
\begin{align}
	v_\mathrm{g}(k_0)
	= \frac{\mathrm{d}\omega_k}{\mathrm{d}k}\bigg|_{k=k_0} = \frac{c^2 k_0}{\sqrt{c^2 k_0^2-\gamma^2}} \ ,
\end{align}
which has the property $v(k_0) > c$. Hence, such a wave packet propagates at least with the velocity $c$ given by Eq.~(\ref{continuum.parameters}).
For waves with $k<\gamma/c$, $z_k$ is purely imaginary and the wave is purely damped. 
Fig.~\ref{fig1}B shows that the continuum approximation (solid curves) captures well the propagation and decay of the waves exhibited by the discrete system (dots).
Note that Eq.~(\ref{telegraph}) is a linear approximation of the nonlinear Eq.~(\ref{eq.phase.model}) in the continuum limit.
Hence, the superposition principle that holds for the solutions to Eq.~(\ref{telegraph}) only holds approximately for solutions of Eq.~(\ref{eq.phase.model}) with long wavelengths.

The interpretation of Eq.~(\ref{telegraph}) as a wave equation enables to determine the range of the damped waves as the distance travelled within the time until decay as
\begin{align}
	R = \frac{c}{\gamma} \propto \varepsilon \sqrt{\kappa \mu} \ .
\end{align}
Hence, the range $R$ increases with increasing coupling strength~$\kappa$ and inertia~$\mu$, while the velocity $c \propto \sqrt{\kappa/\mu}$, Eq.~(\ref{continuum.parameters}) increases with $\kappa$ but decreases with $\mu$. In the limit $\kappa\mu \to \infty$ with a fixed ratio $\kappa/\mu$, we find an infinite range $R$ at finite velocity $c$.
 
\section{Nonlinear approximation: splay states}

We now turn to the full nonlinear approximation Eq.~(\ref{continuum.1d}) with $\lambda \neq 0$.
The only non-damped traveling wave solutions to Eq.~(\ref{continuum.1d}) of the type
\begin{align}
	\Phi(x,t)=\Psi(x-vt) \label{traveling.wave}
\end{align}
 that satisfy the boundary conditions Eqs.~(\ref{bc1}) and (\ref{bc2}) are splay states \cite{Strogatz1993,Choe2010}, 
\begin{align}
	\Psi_m(\xi)=m \xi \ , \qquad v=\frac{\lambda m}{2\gamma} \ , \qquad m\in \mathds{Z} \ , \label{splay.state}
\end{align}
where $\xi=x-vt$. That Eq.~(\ref{splay.state}) provides the only set of non-damped traveling wave solutions to Eq.~(\ref{continuum.1d}) can be seen by inserting the traveling wave ansatz Eq.~(\ref{traveling.wave}) into Eq.~(\ref{continuum.1d}), which leads to
\begin{align}
	(v^2-c^2) \psi' - 2\gamma v \psi - \lambda \psi^2=0
\end{align}
for the derivative $\psi=\Psi'$.  This is a Riccati equation with constant coefficients that can be solved analytically \cite{Bender1999}. The family of all solutions can be obtained as
\begin{align}
	\psi_\alpha(\xi)=\frac{m}{1-\alpha \lambda^{-1} \mathrm{e}^{m \lambda \xi/(v^2-c^2)} } \ ,
\end{align}
where $m = -2\gamma v/\lambda$ and $\alpha \in [-\infty,\infty]$.
The boundary condition Eq.~(\ref{bc2}) translates into $\psi(\xi)=\psi(\xi+\ell)$.
However, the only $\ell$-periodic members of the family $\psi_\alpha$ are the constant solutions $\psi_0=m$ and $\psi_{\infty}=0$. The boundary condition Eq.~(\ref{bc1}) forces $m$ to be integer. Hence, the only family of solutions for $\Psi$ is given by Eq.~(\ref{splay.state}).
Since the second time derivative of the splay states dynamically vanishes, $\partial_t^2 \Phi = v^2 \Psi_m''=0$, they coincide with the splay states found in systems without inertia ($\mu=0$).

To show that Eq.~(\ref{splay.state}) describes the long-wavelength splay states of the discrete model, we use the splay state ansatz $\phi_i(t)=m(\varepsilon i-vt)$ in Eq.~(\ref{eq.phase.model}), which leads to
\begin{align}
\begin{split}
	\frac{2mv}{\kappa} &=\Gamma(\varepsilon m)+\Gamma(- \varepsilon m) \\
	& = \varepsilon^2 m^2 \Gamma''(0) + \mathcal{O}(\varepsilon^4 m^4) \ .
\end{split}
\end{align}
This yields $v \simeq \kappa \varepsilon^2 m \Gamma''(0)/2$, in accordance with Eqs.~(\ref{splay.state}) and (\ref{continuum.parameters}).
Note that this result for the continuum approximation Eq.~(\ref{continuum.1d}) does not rule out the existence of other non-damped traveling wave solutions to Eq.~(\ref{eq.phase.model}) that may arise from higher order nonlinearities in the coupling function $\Gamma$.

\section{Nonlinear approximation: damped waves}

We now investigate how the effects of the nonlinearity in Eq.~(\ref{continuum.1d}) modify the propagation of damped  waves.
We here consider cases where the nonlinearity yields a small perturbation to the solution of the linear telegraph equation (\ref{telegraph}) and use a perturbative expansion in $\lambda$.
Using the ansatz
\begin{align}
	\Phi=\Phi_0 + \lambda  \Phi_1 \label{pt.ansatz}
\end{align}
in Eq.~(\ref{continuum.1d}), we obtain, to zeroth and first order in $\lambda$, the two linear equations
\begin{align}%
	\mathcal{L}\Phi_0 &= 0 \ ,  \label{pt.0} \\%
	\mathcal{L}\Phi_1 &= (\partial_x \Phi_0)^2 \ ,\label{pt.1}%
\end{align}
where the linear operator $\mathcal{L}$ of the telegraph equation was defined in Eq.~(\ref{telegraph}).
In the following, we use Eqs.~(\ref{pt.ansatz}--\ref{pt.1}) to study two generic cases for wave propagation: delocalized waves extending throughout the system and a single localized wave.

\begin{figure*}[t]
\raggedright
\includegraphics[width=14cm]{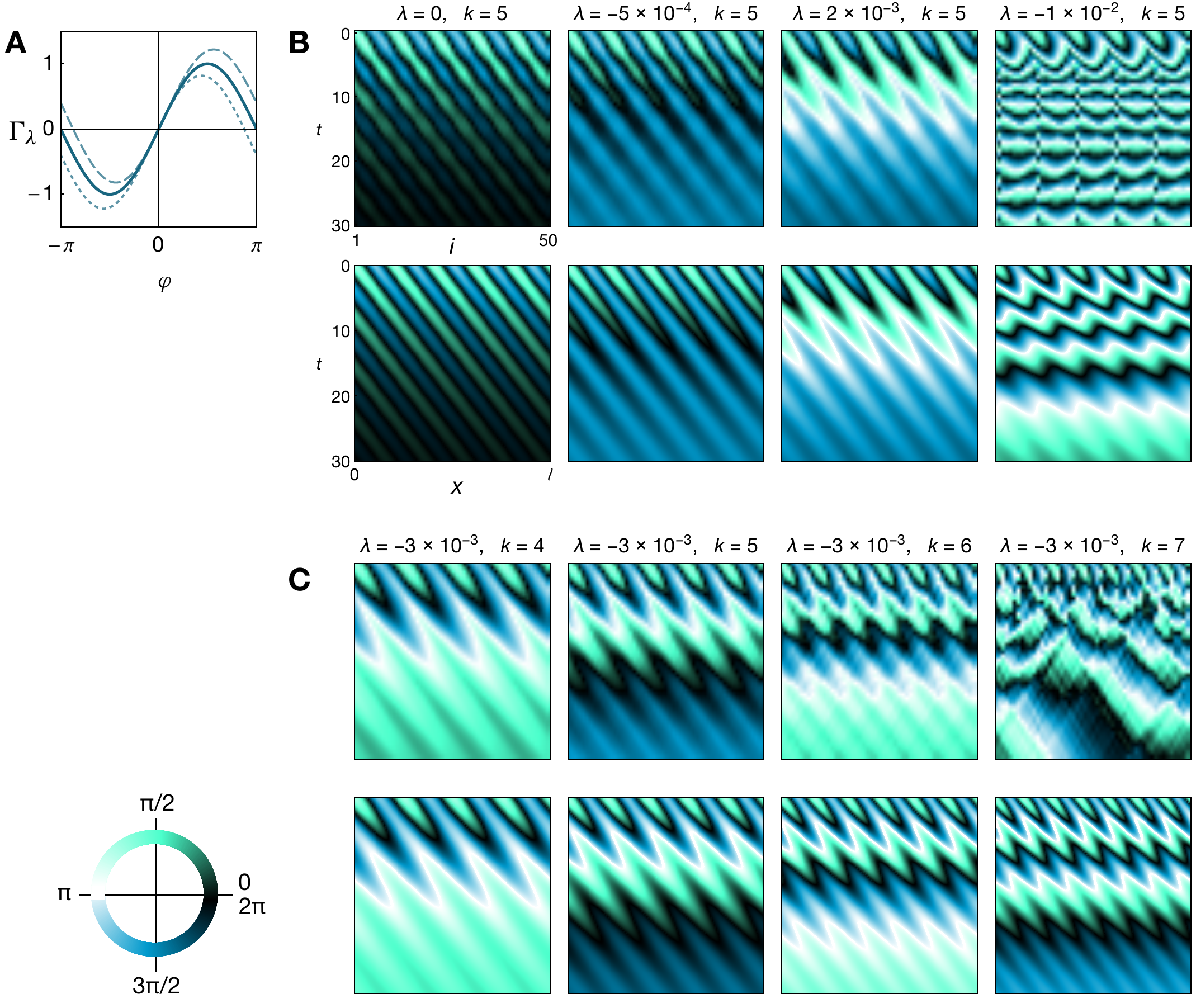}
\caption{Time evolution of delocalized waves in a ring of oscillators. (A) Coupling functions $\Gamma_\lambda$, Eq.~(\ref{coupling.function}), for $\lambda=0$ (solid), $\lambda=-0.2$ (dashed), and $\lambda=0.2$ (dotted). (B) Numerical solutions of the discrete model Eq.~(\ref{eq.phase.model}) (upper row) and the approximate continuum solution (lower row), Eqs.~(\ref{pt.ansatz}), (\ref{damped.wave.solution}) and (\ref{pt.solution}), where the contribution of $\rho(x,t)$ has been neglected, for different $\lambda$ but same wavenumbers $k$.
(C) The same as in B but for fixed $\lambda$ and different wavenumbers $k$.
In B and C, the coupling function of the discrete model is given by Eq.~(\ref{coupling.function}). The other parameters are $\mu=6$, $\kappa=20$, $\Omega=0$ for Eq.~(\ref{eq.phase.model}), implying $c \simeq 0.16$ and $\gamma \simeq 0.08$ for Eq.~(\ref{continuum.1d}). Initial conditions for Eq.~(\ref{eq.phase.model}): $\phi_i(t)=\Phi_0(i\varepsilon,0)$ and $\dot \phi_i(t)=\partial_t\Phi_0(i\varepsilon,0)$, and for Eq.~(\ref{continuum.1d}): $\Phi(x,t)=\Phi_0(x,0)$ and $\partial_t\Phi(x,t)=\partial_t\Phi_0(x,0)$, with $\Phi_0$ given by Eq.~(\ref{damped.wave.solution}) with ${\bar\phi}=3\pi/4$.}%
\label{fig2}%
\end{figure*}%

\subsection{Delocalized waves}

As a first generic scenario we consider the behavior of delocalized waves that extend throughout the system.
As a starting point for the perturbative treatment, we consider the following exact solution to Eq.~(\ref{pt.0}) with boundary conditions (\ref{bc1}) and (\ref{bc2}),
\begin{align}
	\Phi_{0}(x,t)={\bar\phi} \cos(kx-\omega_k t) \mathrm{e}^{-\gamma t} \ , \label{damped.wave.solution}
\end{align}
where $k \in \mathds{Z}$, $\omega_k=\sqrt{k^2 c^2-\gamma^2}$, and ${\bar\phi}$ is the amplitude of the wave.
Eq.~(\ref{damped.wave.solution}) describes a damped plane wave with wavenumber $k$, phase velocity $v_k=\omega_k/k$ and decay rate $\gamma$.
 We only consider cases in which $\omega_k$ is real, that is, $kc>\gamma$.
Eq.~(\ref{pt.1}) can be solved analytically using Fourier transformation, see Appendix~\ref{app:delocalized.solution},
\begin{align}
	\lambda\Phi_1(x,t) &= \big[ 1  -  (1+2 \gamma t) \mathrm{e}^{-2\gamma t} + \rho(x,t) \mathrm{e}^{-\gamma t} \big] \Delta \ , 
\label{pt.solution}
\end{align}
where
\begin{align}
	\Delta &= \frac{\lambda {\bar\phi}^2 k^2}{8 \gamma^2} \ .
\end{align}
 The term $\rho(x,t)$ contains higher harmonics in the wavenumber $k$ but remains bounded as
\begin{align}
	|\rho(x,t)|\leq \frac{2}{\sqrt{3}}+\sqrt{2}\mathrm{e}^{-\gamma t}  \label{pt.bound}
\end{align}
for all cases in which the frequency $\omega_k$ is real, see Appendix~\ref{app:delocalized.solution}.
Since Eq.~(\ref{pt.solution}) implies $\Phi_1 \to \Delta$ as $t \to \infty$, the asymptotic effect of the nonlinearity is a global phase offset of magnitude $\Delta$. Hence, for delocalized waves, the nonlinearity effectively acts as a global source (or sink, depending on the sign of $\lambda$) that transiently contributes to the global frequency.

To demonstrate that our analytical results present an effective description of the discrete system Eq.~(\ref{eq.phase.model}) in the limit of long wavelengths, we compare them to numerical solutions of Eq.~(\ref{eq.phase.model}) for different wavenumbers $k$ and different strengths $\lambda$ of the nonlinearity.
For the discrete model, we choose different coupling functions  from the family
\begin{align}
	\Gamma_\lambda(\varphi)=\sin \varphi- \frac{\lambda}{c^2}(\cos \varphi-1) \ , \label{coupling.function}
\end{align}
where we have identified the parameter $\lambda$ of the continuum approximation according to Eqs.~(\ref{continuum.parameters}), see Fig.~\ref{fig2}A.
Fig.~\ref{fig2}B shows a comparison of solutions to the discrete system Eq.~(\ref{eq.phase.model}) (upper row) and the analytical approximation, Eqs.~(\ref{pt.ansatz}), (\ref{damped.wave.solution}) and (\ref{pt.solution}) (lower row), for different values of $\lambda$.
Compared to the analytical approximation, the discrete model shows undulations in the width of wave peaks which are the result of higher order nonlinearities of the coupling function (first three panels in Fig.~\ref{fig2}B).
For weak nonlinearity (small $\lambda$), the analytical approximation captures the time evolution of the system very well (first three panels in Fig.~\ref{fig2}B), while for stronger nonlinearity (large $\lambda$), the approximation breaks down (rightmost panel in Fig.~\ref{fig2}B) as the discrete system exhibits more complex phase patterns.
Fig.~\ref{fig2}C shows the same comparison for fixed $\lambda$ but different wavenumbers $k$.
For longer initial wavelengths (small $k$), the results of the discrete model and the continuum approximation agree better and better (first three panels in Fig.~\ref{fig2}C).
For small wavelengths (large $k$), the long-wavelength continuum approximation breaks down (rightmost panel in Fig.~\ref{fig2}C).
In all cases considered here, the system eventually reaches the in-phase synchronized state Eq.~(\ref{eq.sync}).

\begin{figure*}[t]
\raggedright
\includegraphics[width=14cm]{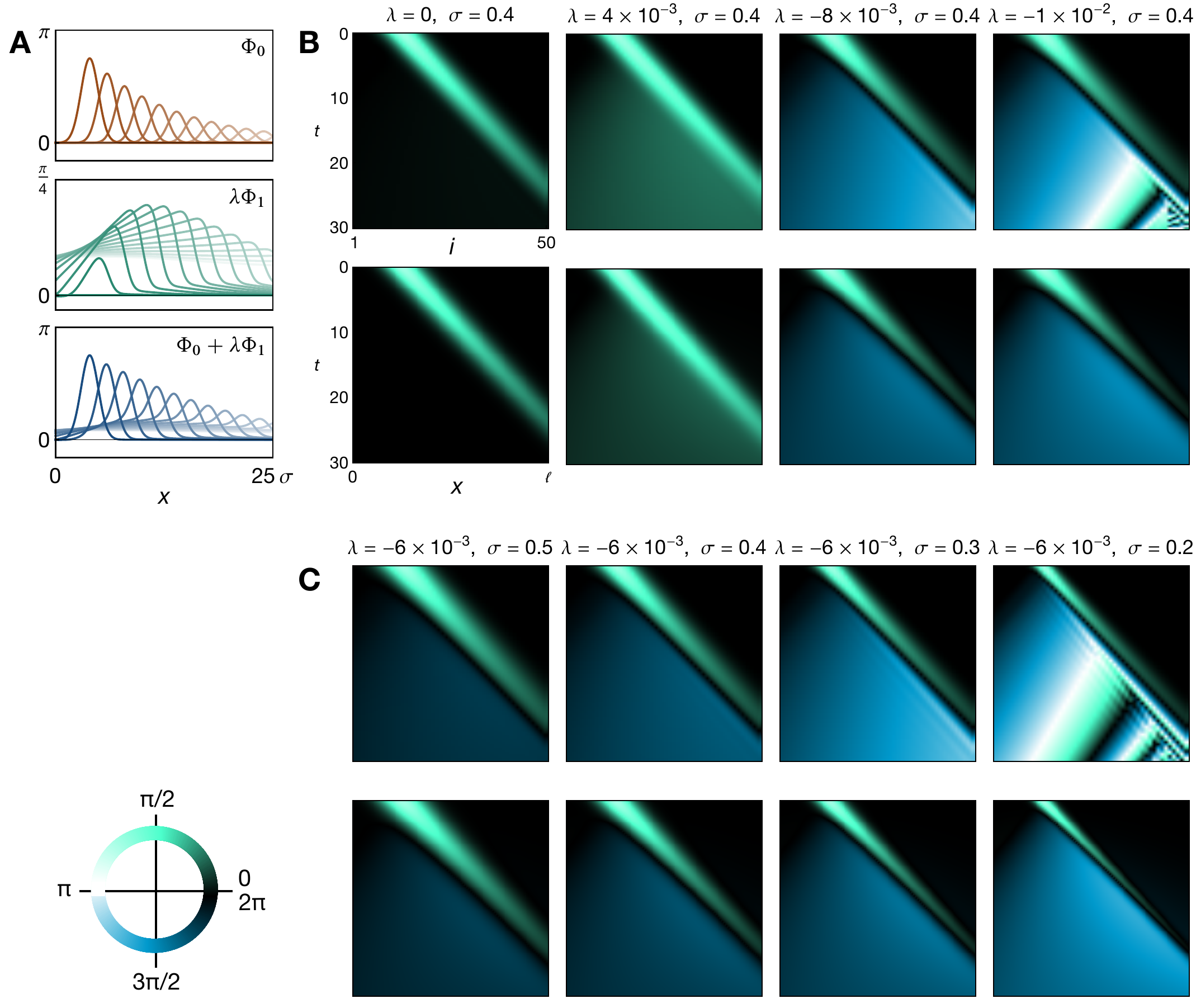}
\caption{Time evolution of a localized wave on an infinite domain. (A) Localized decaying wave $\Phi_0$ (red), Eq.~(\ref{traveling.gaussian}), the perturbative contribution $\lambda \Phi_1$ (green), Eq.~(\ref{damped.perturbation}), and the sum (blue), Eq.~(\ref{pt.ansatz}). Brighter curves correspond to later times ($t=0,{...},60$ in steps of $\Delta t=4$). Parameters as in the second panel from the left of B. (B)
Numerical solutions of the discrete model Eq.~(\ref{eq.phase.model}) (upper row) and the approximate continuum solution (lower row), Eqs.~(\ref{pt.ansatz}), (\ref{traveling.gaussian}) and (\ref{damped.perturbation}), for different $\lambda$ but same width $\sigma$ of the wave.
(C) The same as in B but for fixed $\lambda$ and different widths $\sigma$ of the wave.
The other parameters are $\mu=10$, $\kappa=50$, $\Omega=0$ for Eq.~(\ref{eq.phase.model}), implying $c \simeq 0.2$ and $\gamma = 0.05$ for Eq.~(\ref{continuum.1d}). Initial conditions analogous to Fig.~\ref{fig2}B,C but with $\Phi_0$ given by Eq.~(\ref{traveling.gaussian}) with ${\bar\phi}=3\pi/4$.}
\label{fig3}
\end{figure*}

\subsection{Localized waves}

As a second generic case we consider the propagation of a single localized wave on an infinite spatial domain. Hence, we drop the boundary conditions Eqs.~(\ref{bc1}) and (\ref{bc2}).
As starting point for the perturbative treatment Eqs.~(\ref{pt.ansatz}--\ref{pt.1}), we use a traveling Gaussian wave packet that decays,
\begin{align}
	\Phi_0(x,t)={\bar\phi} \mathrm{e}^{-(x-ct)^2/2\sigma^2}\mathrm{e}^{-\gamma t}  \ , \label{traveling.gaussian}
\end{align}
initially centered around $x=0$ with width $\sigma$, height ${\bar\phi}$, velocity $c$, and decay rate $\gamma$, see Fig.~\ref{fig3}A (red curves).
Eq.~(\ref{traveling.gaussian}) is an approximate solution to Eq.~(\ref{pt.0}) for weak damping, $\mathcal{L}\Phi_0= \mathcal{O}(\gamma^2)$.
Using a Fourier transformation in the spatial domain, an exact solution to Eq.~(\ref{pt.1}) can be obtained for $\hat\Phi_1(k,t)=\int  \Phi_1(x,t) \,  \mathrm{e}^{-\mathrm{i}kx} \, \mathrm{d}x $. However, the corresponding backtransform to real space cannot be represented in terms of elementary functions.
Hence, we approximate the exact solution for $\hat\Phi_1(k,t)$ in terms of Gaussian and trigonometric functions, which permit a backtransformation, see Appendix~\ref{app:localized.solution}. We thus obtain
\begin{align}
\begin{split}
	\Phi_1(x,t)&\simeq {\bar\phi}^2 \big[(1-\mathrm{e}^{-\gamma t})\Pi(x-ct) + \Upsilon(x,t) \\
	&\qquad + \Psi(x+ct)-\Psi(x-ct) \big]\mathrm{e}^{-\gamma t} \ ,
\end{split} \label{damped.perturbation}
\end{align}
where the functions $\Pi$, $\Psi$, and $\Upsilon$ are given by
\begin{align}
	\Pi(x) &= -\frac{1}{4 \gamma c \sigma}{\theta}(x/\sigma) \ ,  \bigg. \label{eq.F} \\
	\Psi(x) &= \frac{1}{16c^2}\big[\mathrm{e}^{-(x/\sigma)^2} + 2(x/\sigma) {\theta} (x/\sigma) \big] \ ,  \bigg. \\
\begin{split}
	\Upsilon(x,t) &= \frac{1-\mathrm{e}^{\gamma t}-\gamma t}{4 \gamma \sigma c}{\theta}(\beta(t) x/\sigma  ) \\
	&\qquad - \beta(t)\frac{\gamma t-\sinh \gamma t}{4 \gamma^2 \sigma^2 } \mathrm{e}^{-(\beta(t) x/\sigma)^2} \ ,  \bigg.
\end{split} \label{eq.L}
\end{align}
with
\begin{align}
	\theta(x)=\int_0^{x} \mathrm{e}^{-y^2} \, \mathrm{d}y
\end{align}
being an error function and 
\begin{align}
	\beta(t)=\frac{1}{\sqrt{1+2c^2 t/\gamma\sigma^2}}
\end{align}
being a time-dependent scaling factor.
Eq.~(\ref{damped.perturbation}) reveals that the effect of the nonlinearity can be understood by three contributions: since $\theta$ essentially is a smoothed step function, the contribution $\Pi(x-ct)$ represents a front with step width $\sigma$ that propagates with the center of the wave. The prefactor $1-\mathrm{e}^{-\gamma t}$ describes a gradual build-up of this front. The behavior of the contributions $\Psi$ and $\Upsilon$ is more complicated but effectively deforms the front into a tail that the traveling wave leaves behind; this behavior of $\Phi_1$ and the full perturbative approximation Eq.~(\ref{pt.ansatz}) is displayed in Fig.~\ref{fig3}A (green and blue curves).
Figs.~\ref{fig3}B,C compare numerical simulations of the discrete model Eq.~(\ref{eq.phase.model}) to this analytical approximation for different $\lambda$ and different widths $\sigma$ of the traveling wave. Again, for weak nonlinearity (small $\lambda$) and long wavelengths (large $\sigma$), the analytical result (\ref{damped.perturbation}) approximates the discrete model very well (first three panels in Figs.~\ref{fig3}B,C). For strong nonlinearity (large $\lambda$) and short wavelengths (small $\sigma$), the continuum approximation breaks down and the higher order nonlinearities in the coupling cause the tail of the wave to evolve into complex phase patterns (rightmost panels in Figs.~\ref{fig3}B,C).

\section{Discussion}

In this work, we have shown that the interplay of  inertia and coupling enables traveling waves in spatially extended oscillator systems, independent of the details of the coupling function.
A spatial continuum approximation for long wavelengths has been derived that elucidates the mechanism of wave propagation and permits the analytical computation of key observables such as the wave speed and damping rate.
We found that the range and velocity of these waves is governed by the magnitude of inertia and coupling strength.
For the case of purely sinusoidal coupling, wave propagation can be described by a linear telegraph equation to a very good approximation.
For general coupling functions, the continuum approximation is nonlinear.
This nonlinearity modifies the behavior of damped waves and enables splay states with finite frequencies corresponding to persistent traveling waves.
For delocalized waves, we found that the nonlinearity transiently modifies the global frequency of the system; for localized waves, the nonlinearity leads to a tail that the traveling wave packet transiently leaves behind.

The continuum approximation Eq.~(\ref{continuum.corotating})  is a generalization of many well-known equations.
It generalizes the telegraph equation (\ref{telegraph}) describing damped linear waves to include nonlinear effects.
It is a generalization of the Kardar--Parisi--Zhang (KPZ) equation\cite{Kardar1986} without noise to include a second time derivative.
Differentiating Eq.~(\ref{continuum.1d}) with respect to $x$ and defining $u=\partial_x\Phi$ yields a generalized Burgers' equation $\partial_t^2 u + 2 \gamma \partial_t u = c^2 \partial_x^2 u +\lambda u \partial_x u$, which contains an inertial term.
For $\lambda \neq 0$ and appropriate boundary conditions, this generalized Burgers' equation is known to exhibit non-decaying traveling wave solutions \cite{Kar2003}.
Moreover, equations of the type Eq.~(\ref{continuum.corotating}) have been used to describe the propagation of electromechanical waves in large electric power systems\cite{Thorp1998,Xu2014}.

The results presented here show that the interplay of coupling and inertia (i.e., slow frequency adaption) enables signal transmission and the distribution of spatial information through phase waves.
These findings may therefore be relevant for natural and engineered coupled oscillator systems with slow frequency adaptation in which phase patterns can play a functional or interfering role, such as in biological oscillators \cite{Ermentrout1991}, power grids \cite{Filatrella2008,Rohden2012}, and systems of coupled electronic clocks \cite{Pollakis2014}.
More generally, the present work highlights how the internal dynamics of an oscillator can crucially contribute to the emergence of collective phenomena that appear in coupled systems.
While we here investigated phase waves with long wavelengths, more general phase dynamics will unfold if the long-wavelength criterion is abandoned.
This remains a rich subject for future research.

\section*{Acknowledgements}

I gratefully acknowledge useful discussions with Lucas Wetzel, Luis Morelli, Frank J\"ulicher, Pablo Sartori, and Julia Kynast.
I also thank the Center for Advancing Electronics Dresden cfAED for providing an inspiring collaborative atmosphere.

\begin{appendix}

\section{Linear stability analysis of the in-phase synchronized state}
\label{app:stability}

To assess under which circumstances the in-phase synchronized state Eq.~(\ref{eq.sync}) is stable, we use the ansatz $\phi_i(t)=\Omega t+\eta \zeta_i(t)$ in Eq.~(\ref{eq.phase.model}), where $\zeta_i$ is a perturbation of order unity and $\eta$ is a small expansion parameter. Linearizing the resulting dynamics in $\eta$ yields the governing equations for the perturbations as
\begin{align}
	\mu \ddot \zeta_i + \dot\zeta_i = \kappa \Gamma'(0) \sum_{j=1}^N \tilde w_{ij} (\zeta_j-\zeta_i) \ ,
\end{align}
where $\tilde w_{ij}=w_{ij}/n_i$ is the normalized adjacency matrix.
To decouple this set of equations, we define the collective modes $\vartheta_i = \sum_j (m^{-1})_{ij} \zeta_j$, where $(m_{ij})$ is the matrix that diagonalizes $(\tilde w_{ij})$ according to $\sum_{kl} (m^{-1})_{ik} \tilde w_{kl} m_{kj} = \mathrm{diag}(e_1,{...},e_N)$ and $e_i$ are the eigenvalues of $\tilde w$. This yields the decoupled set of equations $\mu \ddot\vartheta_i + \dot\vartheta_i + g_i \vartheta_i=0$, where $g_i= \kappa \Gamma'(0)  (1-e_i)$. The exponential ansatz $\vartheta_i(t)=\mathrm{e}^{z_i t}$ yields the characteristic equation $\mu z_i^2 +  z_i + g_i = 0$ with solutions
\begin{align}
	z_{i,\pm} = \frac{-1 \pm \sqrt{1-4 \mu g_i}}{2\mu} \ .
\end{align}
The in-phase synchronized state is linearly stable if and only if $\mathrm{Re}\,z_{i,\pm}<0$ for all $i$. Using Gershgorin's circle theorem, it has been shown \cite{EarlStrogatz2003} that the eigenvalues $e_i$ of the normalized adjacency matrix $\tilde w_{ij}$ satisfy $|e_i| \leq 1$. This implies $\mathrm{sign}\,g_i = \mathrm{sign}\,\kappa \Gamma'(0)$. Consequently, $\mathrm{Re}\, z_{i,\pm}<0$ for all modes $i$ if and only if $\kappa \Gamma'(0)>0$, in which case the synchronized state is linearly stable. This is the same criterion as for systems without inertia \cite{EarlStrogatz2003}.

\section{Derivation of the spatial continuum approximation Eq.~(\ref{continuum.corotating})}
\label{app:continuum}
\noindent%
In this section, we derive the spatial continuum approximation Eq.~(\ref{continuum.corotating}) from the discrete model Eq.~(\ref{eq.phase.model}) with nearest-neighbor coupling. We proceed along the lines of Refs.~\cite{Kuramoto1984b,Ares2012}. We first show the derivation for $d=1$ dimension and then extend the result to arbitrary dimensionality $d$.
The one-dimensional nearest-neighbor adjacency matrix is given by $w_{ij}=\delta_{i,j-1}+\delta_{i,j+1}$.
We replace $\phi_i(t)$ by $\Phi(x,t)$ and $\phi_{i\pm \sigma}(t)$ by $\Phi(x\pm \sigma \varepsilon,t)$ with $\sigma \in \{-1,1\}$ in Eq.~(1), where $\varepsilon$ is the lattice spacing,
\begin{align}
	\mu \partial_t^2 \Phi + \partial_t \Phi = \Omega+ \frac{\kappa}{2} \sum_{\sigma=\pm1} \Gamma(\Phi(x + \sigma \varepsilon,t)-\Phi(x,t)) \ , \bigg. \label{continuum.portation}
\end{align}
and expand the coupling function in the lattice spacing $\varepsilon$,
\begin{widetext}
\begin{align}
\begin{split}
	&\Gamma(\Phi(x + \sigma \varepsilon,t)-\Phi(x,t))
	= \sigma\varepsilon \Gamma'(0) \partial_x \Phi(x,t)  + \frac{\varepsilon^2}{2} \Gamma'(0) \partial_x^2 \Phi(x,t) + \frac{\varepsilon^2}{2} \Gamma''(0) [\partial_x \Phi(x,t)]^2 + \mathcal{O}(\varepsilon^3) \ ,
\end{split} \label{expansion}
\end{align}
where we have used $\sigma^2=1$. Using the expansion Eq.~(\ref{expansion}) in Eq.~(\ref{continuum.portation}), we find that terms of linear order in $\varepsilon$ cancel as the $\sigma$-sum runs over the values $1$ and $-1$. Dividing Eq.~(\ref{continuum.portation}) by $\mu$ and dropping terms of order $\varepsilon^3$, we obtain
\begin{align}
	\partial_t^2 \Phi + 2\gamma \partial_t \Phi = \tilde\Omega + c^2 \partial_x^2 \Phi + \lambda (\partial_x\Phi)^2 \ , 
\end{align}
where $\tilde\Omega=\Omega/\mu$, and $\gamma$, $c$, and $\lambda$ have been defined in Eqs.~(\ref{continuum.parameters}).
The analogue of Eq.~(\ref{continuum.portation}) for arbitrary dimension $d$ is
\begin{align}
	\mu \partial_t^2 \Phi + \partial_t \Phi = \Omega + \frac{\kappa}{2d} \sum_{n=1}^d \sum_{\sigma=\pm 1} \Gamma(\Phi(\mathbf{x}+ \sigma \varepsilon \mathbf{e}_n,t)-\Phi(\mathbf{x},t)) \ , \label{continuum.portation.md}
\end{align}
where $\mathbf{e}_n$ is the unit vector in $n$-direction. The corresponding expansion of the coupling function is given by
\begin{align}
\begin{split}
	&\Gamma(\Phi(\mathbf{x}+ \sigma \varepsilon \mathbf{e}_n,t)-\Phi(\mathbf{x},t))
	= \sigma\varepsilon \Gamma'(0) \frac{\partial \Phi}{\partial x_n}(\mathbf{x},t)  + \frac{\varepsilon^2}{2} \Gamma'(0) \frac{\partial^2 \Phi}{\partial x_n^2}(\mathbf{x},t) + \frac{\varepsilon^2}{2} \Gamma''(0) \bigg(\frac{\partial \Phi}{\partial x_n}(\mathbf{x},t)\bigg)^2 + \mathcal{O}(\varepsilon^3) \ .
\end{split} \label{expansion.md}
\end{align}
\end{widetext}
Dividing Eq.~(\ref{continuum.portation.md}) by $\mu$ and using the expansion Eq.~(\ref{expansion.md}) yields
\begin{align}
\begin{split}
	\partial_t^2 \Phi + 2\gamma \partial_t \Phi &= \tilde\Omega +  \sum_{n=1}^d \bigg[ c^2 \frac{\partial^2 \Phi}{\partial x_n^2} + \lambda \left(\frac{\partial \Phi}{\partial x_n} \right)^2 \bigg] \\
	&= \tilde\Omega + c^2 \nabla^2 \Phi + \lambda (\nabla\Phi)^2 \ . \bigg.
\end{split}
\end{align}
This completes the derivation of Eq.~(\ref{continuum.corotating}).

\section{Derivation of Eqs.~(\ref{pt.solution}) and (\ref{pt.bound})}
\label{app:delocalized.solution}

\noindent%
In this section, we derive Eq.~(\ref{pt.solution}), that is, we solve Eq.~(\ref{pt.1}) with $\Phi_0$ given by Eq.~(\ref{damped.wave.solution}). For convenience, we here set the amplitude $\bar\phi=1$ and restore it later. The rhs of Eq.~(\ref{pt.1}), given by $(\partial_x \Phi_0)^2 = k^2 \sin(kx-\omega_k t)^2 \mathrm{e}^{-2\gamma t}$ with $\omega_k=\sqrt{k^2 c^2-\gamma^2}$, can be rewritten using standard trigonometric relations,
\begin{align}
\begin{split}
	(\partial_x \Phi_0)^2
	&= \frac{k^2}{2} \bigg[ 1 - \cos(2kx) \cos(2\omega_k t) \\
	&\qquad\qquad - \sin(2kx) \sin(2\omega_k t) \bigg] \mathrm{e}^{-2\gamma t} \ .
\end{split}
\end{align}
Hence, we make the ansatz $\Phi_1(x,t)=\varphi_0(t) + \cos(2kx)\varphi_1(t) +\sin(2kx) \varphi_2(t)$ in Eq.~(\ref{pt.1}), which yields the following ordinary differential equations for $\varphi_0$, $\varphi_1$, and $\varphi_2$,
\begin{align}
\begin{split}
	\ddot \varphi_0 + 2 \gamma \dot \varphi_0 &= \frac{k^2}{2}\mathrm{e}^{-2\gamma t} \ , \\
	\ddot \varphi_1 + 4k^2 c^2 \varphi_1 + 2 \gamma \dot \varphi_1 &= -\frac{k^2}{2} \cos(2\omega_k t) \mathrm{e}^{-2\gamma t} \ , \\
	\ddot \varphi_2 + 4k^2 c^2 \varphi_2 + 2 \gamma \dot \varphi_2 &= -\frac{k^2}{2} \sin(2\omega_k t) \mathrm{e}^{-2\gamma t} \ .
\end{split} \label{ft.reduced.odes}
\end{align}
Solving Eqs.~(\ref{ft.reduced.odes}) with the initial conditions $\varphi_0(0)=\varphi_1(0)=\varphi_2(0)=0$ leads to Eq.~(\ref{pt.solution}), where $\rho(x,t)$ is given by
\begin{widetext}
\begin{align}
\begin{split}
	\frac{1}{\Lambda_k^2}\rho(x,t) &= \bigg(\frac{\omega_k}{\omega_{2k}} \sin(\omega_{2k} t)+ \frac{\omega_k}{\gamma} \cos(\omega_{2k} t) \bigg) \sin(2kx)  - \bigg(  \frac{2 k^2 c^2-\gamma^2}{\gamma \omega_{2k}} \sin(\omega_{2k} t)-\cos(\omega_{2k} t) \bigg) \cos(2kx) \\
	&\quad  - \bigg[ \bigg( \sin(2 \omega_k t)+\frac{\omega_k}{\gamma} \cos(2\omega_k t) \bigg) \sin(2kx)  - \bigg( \frac{\omega_k}{\gamma} \sin(2 \omega_k t)- \cos(2\omega_k t)  \bigg)  \cos(2kx) \bigg] \mathrm{e}^{-\gamma t}\ .
\end{split}
\end{align}
\end{widetext}
with $\Lambda_k=\gamma/kc$. We now prove the bound (\ref{pt.bound}).
Using the triangle inequality $|a+b| \leq |a|+|b|$ and the trigonometric inequality $|a \cos x + b \sin x| \leq \sqrt{a^2+b^2}$, we successively obtain the bounds
\begin{widetext}
\begin{align*}
	\frac{1}{\Lambda_k^2}|\rho(x,t)| &\leq \bigg| \bigg(\frac{\omega_k}{\omega_{2k}} \sin(\omega_{2k} t)+ \frac{\omega_k}{\gamma} \cos(\omega_{2k} t) \bigg) \sin(2kx)
	- \bigg(  \frac{2 k^2 c^2-\gamma^2}{\gamma\omega_{2k}} \sin(\omega_{2k} t) -\cos(\omega_{2k} t) \bigg) \cos(2kx) \bigg| \\
	&\quad  + \bigg| \bigg( \sin(2 \omega_k t)+\frac{\omega_k}{\gamma} \cos(2\omega_k t) \bigg) \sin(2kx)   - \bigg( \frac{\omega_k}{\gamma} \sin(2 \omega_k t)- \cos(2\omega_k t)  \bigg)  \cos(2kx) \bigg| \mathrm{e}^{-\gamma t} \\
	&\leq \bigg[ \bigg(\frac{\omega_k}{\omega_{2k}} \sin(\omega_{2k} t)+ \frac{\omega_k}{\gamma} \cos(\omega_{2k} t) \bigg)^2 
	+ \bigg(  \frac{2 k^2 c^2-\gamma^2}{\gamma\omega_{2k}} \sin(\omega_{2k} t)-\cos(\omega_{2k} t) \bigg)^2  \bigg]^{1/2} \\
	&\quad  + \bigg[ \bigg( \sin(2 \omega_k t)+\frac{\omega_k}{\gamma} \cos(2\omega_k t) \bigg)^2  + \bigg( \frac{\omega_k}{\gamma} \sin(2 \omega_k t)- \cos(2\omega_k t)  \bigg)^2 \bigg]^{1/2} \mathrm{e}^{-\gamma t} \\
	&\leq \bigg[ \bigg(\frac{\omega_k}{\omega_{2k}} \bigg)^2+ \bigg( \frac{\omega_k}{\gamma} \bigg)^2 + \bigg( \frac{2 k^2 c^2-\gamma^2}{\gamma\omega_{2k}} \bigg)^2+ 1  \bigg]^{1/2}   + \bigg[ 2+ 2\bigg(\frac{\omega_k}{\gamma} \bigg)^2  \bigg]^{1/2} \mathrm{e}^{-\gamma t} \\
	&= \frac{\sqrt{2}}{|\Lambda_k|} \left( \sqrt{\frac{1-\Lambda_k^2/2}{1-\Lambda_k^2/4}} + \mathrm{e}^{-\gamma t} \right) \ .
\end{align*}
\end{widetext}
Hence, $|\rho(x,t)| \leq \sqrt{2}  b(\Lambda_k) + \sqrt{2} |\Lambda_k| \mathrm{e}^{-\gamma t}$, where
\begin{align}
	b(\Lambda) =|\Lambda| \sqrt{\frac{1-\Lambda^2/2}{1-\Lambda^2/4}} \ .
\end{align}
We only consider cases in which the wave is not purely damped, which corresponds to $\smash{\omega_k=\sqrt{k^2c^2-\gamma^2}}$ being real. This implies $|k c| \geq |\gamma|$ or, equivalently, $|\Lambda_k| \leq 1$. The bound (\ref{pt.bound}) then follows from the observation that $\smash{0 < b(\Lambda) \leq \sqrt{2/3}}$ for $|\Lambda| \leq 1$.

\section{Derivation of Eq.~(\ref{damped.perturbation})}
\label{app:localized.solution}

\noindent%
In this section, we derive the approximate solution Eq.~(\ref{damped.perturbation}).
We start from Eq.~(\ref{pt.1}) with $\Phi_0$ given by Eq.~(\ref{traveling.gaussian}). As in Appendix~\ref{app:delocalized.solution}, we set the amplitude $\bar\phi=1$ for convenience and restore it later.
Expressing $\Phi_1$ by its spatial Fourier transform as $\Phi_1(x,t)=(2\pi)^{-1} \int \mathrm{e}^{\mathrm{i}kx} \hat{\Phi}_1(k,t) \, \mathrm{d}k$, Eq.~(\ref{pt.1}) can be written as
\begin{align}
\begin{split}
	&\partial_t^2 \hat{\Phi}_1 + k^2 c^2 \hat{\Phi}_1 + 2\gamma \partial_t \hat{\Phi}_1 = \int \mathrm{d}x \, \mathrm{e}^{-\mathrm{i}kx} (\partial_x\Phi_0)^2 \\
	&\qquad = \frac{\sqrt{\pi}}{\sigma} \left( \frac{1}{2}- \frac{k^2 \sigma^2}{4} \right) \mathrm{e}^{-k^2\sigma^2/4} \mathrm{e}^{-(2\gamma+\mathrm{i}kc)t} \ .
\end{split} \label{sft}
\end{align}
Eq.~(\ref{sft}) is a linear ordinary differential equation in $t$ whose exact solution is given by
\begin{align}
\begin{split}
	\hat{\Phi}_1(k,t) &=\frac{\sqrt{\pi}}{2 \gamma^2 \sigma}  \bigg( (y_k - \mathrm{i}  )   U_{\gamma t}(y_k) \gamma t + \mathrm{i} V_{\gamma t}(y_k) \\
	&\qquad\qquad -\mathrm{i} \mathrm{e}^{-(\mathrm{i}y_k+1)\gamma t}  \bigg) \frac{W_{\gamma\sigma/2c}(y_k)}{y_k} \mathrm{e}^{-\gamma t} \ , 
\end{split}\label{ft.exact}
\end{align}
where $y_k = kc/\gamma$ and the functions $W_\alpha$, $U_\alpha$, and $V_\alpha$ are given by
\begin{align}
	W_\alpha(y) &=\left(\frac{1}{2}-\alpha^2 y^2 \right) \mathrm{e}^{-\alpha^2 y^2} \label{W.exact}
\end{align}
and
\begin{align}
	U_\alpha(y) &= \frac{\mathrm{e}^{\alpha\sqrt{1-y^2}}-\mathrm{e}^{-\alpha\sqrt{1-y^2}}}{2\alpha \sqrt{1-y^2}} \ , \label{U.exact} \\
	V_\alpha(y) &= \frac{\mathrm{e}^{\alpha\sqrt{1-y^2}}+\mathrm{e}^{-\alpha\sqrt{1-y^2}}}{2} \ . \label{V.exact}
\end{align}

\paragraph{Approximation of $W_\alpha$}

The function $W_\alpha$, Eq.~(\ref{W.exact}), can be approximated by $W_\alpha \simeq \tilde W_\alpha$, where
\begin{align}
	\tilde W_\alpha(y) &=\frac{1}{2} \mathrm{e}^{-\alpha^2 y^2} \ , \label{W.tilde}
\end{align}
since the term $-\alpha^2 y^2$ becomes of the same magnitude as the term $1/2$ at $y=1/\sqrt{2}\alpha$; this is of the same order as the decay length $1/\alpha$ of the Gaussian term $\smash{\mathrm{e}^{-\alpha^2 y^2}}$, which thus suppresses this contribution. The maximum distance between $W_\alpha$ and $\tilde W_\alpha$ is given by $D_W=\sup_y | W_\alpha(y)-\tilde W_\alpha(y)|=\alpha \mathrm{e}^{-1}/\sqrt{2}$, which decreases with decreasing $\alpha$. Since $\alpha \propto \sigma$ in Eq.~(\ref{ft.exact}), the approximation becomes more accurate for sharper localized phase perturbations.

\paragraph{Approximation of $U_\alpha$ and $V_\alpha$}

Due to the expression $\sqrt{1-y^2}$ in the exponentials, the functions $U_\alpha$ and $V_\alpha$ do not possess a Fourier transform in terms of elementary functions. Hence, we approximate $U_\alpha$ and $V_\alpha$ by Gaussian and trigonometric functions, which permit a backtransform of $\smash{\hat{\Phi}_1}$ to real space.
We construct approximating functions in the following way. First, note that, while $\smash{\sqrt{1-y^2}}$ may become imaginary for $|y|>1$, both $U_\alpha$ and $V_\alpha$ are real. For $|y|\leq 1$, this is obvious, while for $|y|>1$, we can rewrite $U_\alpha$ and $V_\alpha$ as
\begin{align}
\begin{split}
	U_\alpha(y)\Big|_{|y|>1} &= \frac{\mathrm{e}^{\mathrm{i} \alpha\sqrt{y^2-1}}-\mathrm{e}^{-\mathrm{i} \alpha\sqrt{y^2-1}}}{2\mathrm{i} \alpha\sqrt{y^2-1}} = \frac{\sin( \alpha \sqrt{y^2-1})}{\alpha\sqrt{y^2-1}} \ , \\
	V_\alpha(y)\Big|_{|y|>1} &= \frac{\mathrm{e}^{\mathrm{i} \alpha\sqrt{y^2-1}}+\mathrm{e}^{-\mathrm{i} \alpha\sqrt{y^2-1}}}{2}=\cos( \alpha \sqrt{y^2-1}) \ .
\end{split}
\end{align}
For large $|y|$, we approximate $\sqrt{y^2-1} \simeq |y|$ and thus obtain the asymptotic behavior
\begin{align}
\begin{split}
	U_\alpha(y)\Big|_{|y|\gg1} &\simeq \frac{\sin \alpha |y|}{\alpha|y|}=\frac{\sin \alpha y}{\alpha y} \ , \\
	V_\alpha(y)\Big|_{|y|\gg1} &\simeq \cos( \alpha y) \ .
\end{split} \label{asymptotic}
\end{align}
\begin{figure*}[t]
\centering
\includegraphics[width=15cm]{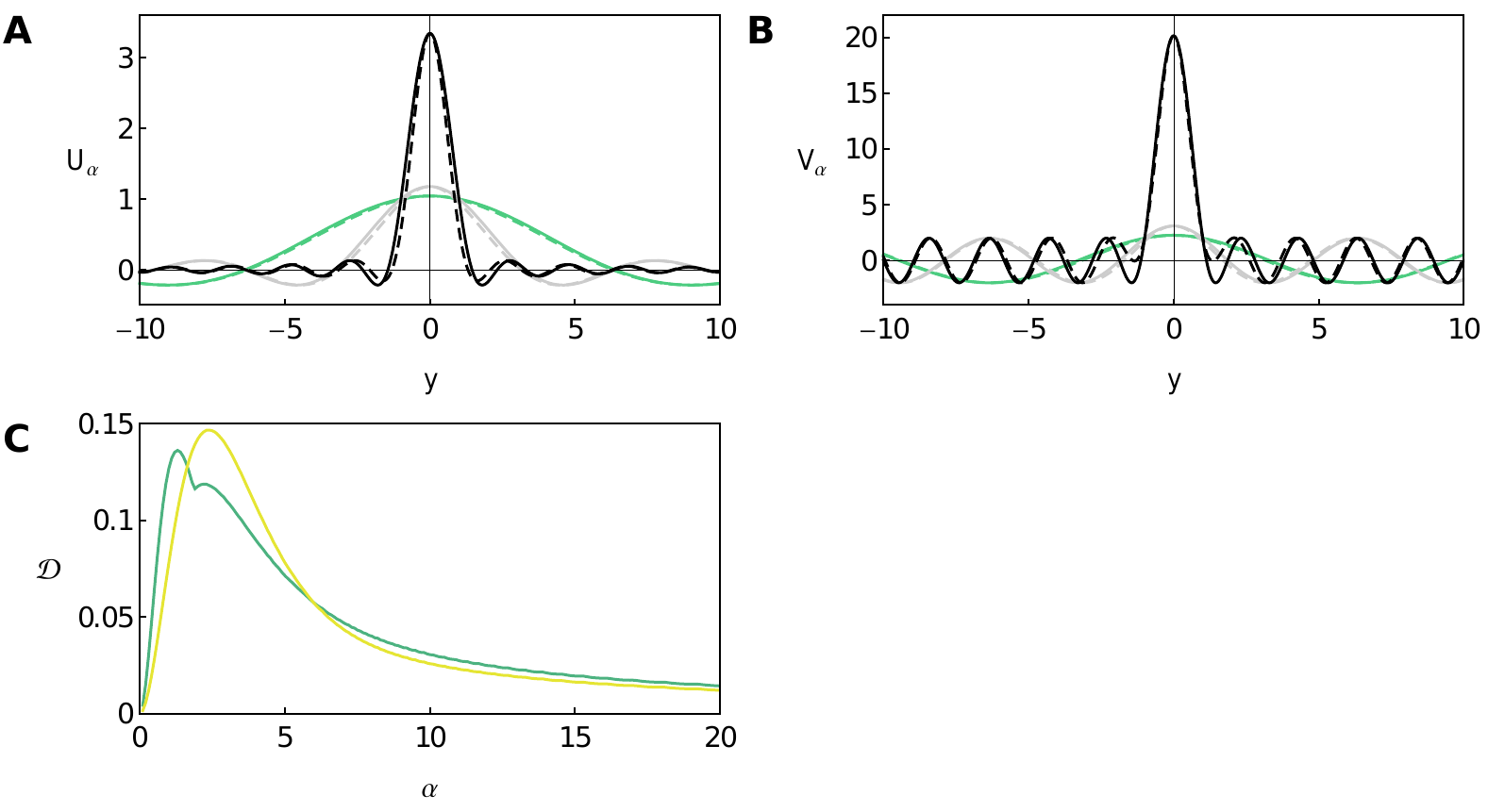}
\caption{(A,B) Comparison of the functions $U_\alpha$ and $V_\alpha$ (solid curves), Eqs.~(\ref{U.exact}) and (\ref{V.exact}), with the approximations $\tilde U_\alpha$ and $\tilde V_\alpha$ (dashed curves), Eqs.~(\ref{tilde}), for $\alpha=0.5$ (green), $\alpha=1$ (gray), and $\alpha=2$ (black). (C) Relative distances $\mathcal{D}_U$ (yellow) and $\mathcal{D}_V$ (green), Eqs.~(\ref{error}), as a function of $\alpha$.}
\label{figS1}
\end{figure*}%
Both $U_\alpha$ and $V_\alpha$ exhibit a global maximum at $y=0$, see Fig.~\ref{figS1}A,B.
For small $|y|$, we approximate this feature of $U_\alpha$ and $V_\alpha$ by Gaussian functions. To obtain approximations consistent with the asympotic behavior given by Eqs.~(\ref{asymptotic}), we make the ansatz
\begin{align}
\begin{split}
	U_\alpha(y) &= \mathrm{e}^{u_\alpha(y)} + \frac{\sin \alpha y}{\alpha y} \ , \\
	V_\alpha(y) &= \mathrm{e}^{v_\alpha(y)} + \cos  \alpha y \ , 
\end{split} \label{ansatz}
\end{align} 
where the functions $u_\alpha$ and $v_\alpha$ are to be determined in such a way that they capture the behavior of $U_\alpha$ and $V_\alpha$ for small $|y|$.
To this end, we solve Eqs.~(\ref{ansatz}) for $u_\alpha$ and $v_\alpha$ and, using (\ref{U.exact}) and (\ref{V.exact}), expand the resulting expressions in $y$,
\begin{align}
\begin{split}
	u_\alpha(y) &= \log \bigg( U_\alpha(y) - \frac{\sin \alpha y}{\alpha y} \bigg) \\
	&= \log \bigg( \frac{\sinh \alpha}{\alpha}-1 \bigg)  -\frac{p_\alpha}{2} y^2 + \mathcal{O}(y^3) \ , 
\end{split} \label{p.expansion} \\[6pt]
\begin{split}
	v_\alpha(y) &= \log ( V_\alpha(y) - \cos  \alpha y  ) \\
	&= \log(\cosh\alpha-1) -  \frac{q_\alpha}{2}  y^2 + \mathcal{O}(y^3)  \ , 
\end{split}\label{q.expansion}
\end{align}
where
\begin{align}
\begin{split}
	p_\alpha &= \frac{\alpha+\alpha^3/3-\alpha\cosh\alpha}{\alpha-\sinh\alpha}-1 \ , \\
	q_\alpha &= \frac{1}{2}\frac{\alpha \sinh \alpha-\alpha^2}{\sinh(\alpha/2)^2} \ .
\end{split}
\end{align}
Hence, we obtain the approximations $U_\alpha \simeq U^*_\alpha$ and $V_\alpha \simeq V^*_\alpha$, where
\begin{align}
\begin{split}
	 U^*_\alpha(y) &= \left( \frac{\sinh\alpha}{\alpha}-1 \right) \mathrm{e}^{-p_\alpha y^2/2} + \frac{\sin \alpha y}{\alpha y} \ , \\
	 V^*_\alpha(y) &=  (\cosh\alpha-1)  \mathrm{e}^{-q_\alpha y^2/2}  +  \cos \alpha y \ .
\end{split}
	 \label{q.tilde}
\end{align}
Note that $p_\alpha$ and $q_\alpha$ are complicated functions of $\alpha$ which lead to a complicated time dependence of the approximation for Eq.~(\ref{ft.exact}), where $\alpha=\gamma t$.
We can considerably simplify the approximations for $U_\alpha$ and $V_\alpha$ by further approximating $\smash{U^*_\alpha \simeq \tilde U_\alpha}$ and $\smash{V^*_\alpha \simeq \tilde V_\alpha}$, where
\begin{align}
\begin{split}
	\tilde U_\alpha(y) &= \left( \frac{\sinh\alpha}{\alpha}-1 \right) \mathrm{e}^{-\alpha y^2/2} + \frac{\sin \alpha y}{\alpha y} \ , \\
	\tilde V_\alpha(y) &=  (\cosh\alpha-1)  \mathrm{e}^{-\alpha y^2/2}  + \cos \alpha y \ .
\end{split}
	\label{tilde}
\end{align}
Justifying the approximations (\ref{tilde}) is not straightforward as they do not solely rely on an approximation of $p_\alpha$ and $q_\alpha$. Hence, we first give some intuitive reasoning and afterwards show that the approximations are valid by numerically computing the relative distance between the approximations (\ref{tilde}) and the exact functions (\ref{U.exact}) and (\ref{V.exact}).
The approximations (\ref{tilde}) are based on two observations:
(i) for $\alpha\to \infty$, $p_\alpha$ and $q_\alpha$ asymptotically behave as $p_\alpha \sim \alpha-1$ and $q_\alpha \sim \alpha$, which is straighforward to show, and (ii) the prefactors $(\alpha^{-1}\sinh\alpha-1)=\alpha^2/6+\mathcal{O}(\alpha^4)$ and $(\cosh\alpha-1)=\alpha^2/2+\mathcal{O}(\alpha^4)$ in Eqs.~(\ref{tilde}) suppress the effects of the Gaussian terms for small $\alpha$ and eventually vanish for $\alpha \to 0$.
Hence, our approximation strategy is to replace $p_\alpha$ and $q_\alpha$ by their asymptotic behavior at large $\alpha$.
The effects of the large error for small $\alpha$ that is thus introduced is then suppressed by the prefactors.
While the replacement of $q_\alpha$ by $\alpha$ leads to no difficulties at this stage, the replacement of $p_\alpha$ by $\alpha-1$ would lead to positive exponents in the Gaussian $\smash{\mathrm{e}^{-(\alpha-1)y^2}}$ for $\alpha<1$, which would consequently lead to an unbounded growth as $|y| \to \infty$.
Hence, we approximate $\alpha-1\simeq \alpha$, which is a reasonable approximation for $\alpha \gg 1$, while still leading to a bounded behavior of $\smash{\tilde U_\alpha}$ for $\alpha<1$. Again, the effects of the large error introduced for small $\alpha$ remains suppressed by the prefactor.
Note that the thus introduced deviations from the exact values $p_\alpha$ and $q_\alpha$ only alter the width of Gaussians and not their height; for instance, the exact identities $\smash{\tilde U_\alpha(0)}=U_\alpha(0)$ and $\smash{\tilde V_\alpha(0)}=V_\alpha(0)$ are preserved by this approximation. Fig.~\ref{figS1}A,B show comparisons of the exact function $U_\alpha$ and $V_\alpha$, Eqs.~(\ref{U.exact}) and (\ref{V.exact}), with the approximations $\tilde U_\alpha$ and $\tilde V_\alpha$, Eqs.~(\ref{tilde}), for different values of $\alpha$.
To show that that Eqs.~(\ref{tilde}) provide reasonable approximations, we numerically compute the relative distances
\begin{align}
\begin{split}
	\mathcal{D}_U(\alpha)  &= \frac{\| U_\alpha-\tilde U_\alpha \|_\infty}{\| U_\alpha \|_\infty} \ , \\
	\mathcal{D}_V(\alpha)  &= \frac{\| V_\alpha-\tilde V_\alpha \|_\infty}{\| V_\alpha \|_\infty} \ ,
\end{split}
\label{error}
\end{align}
where $\| f \|_\infty = \sup_y |f(y)|$ is the uniform norm. The result is shown in Fig.~\ref{figS1}C highlighting that the approximations (\ref{tilde}) work especially well for very small and very large $\alpha$, while satisfying $\mathcal{D}_U < 15\%$ and $\mathcal{D}_V < 15\%$ for all $\alpha$ in the computed range. 
In summary, the approximation of $\hat\Phi_1$ is given by
\begin{align}
\begin{split}
	\hat{\Phi}_1(k,t) &\simeq \frac{\sqrt{\pi}}{2 \gamma^2 \sigma}  \bigg( (y_k - \mathrm{i}  )  \tilde U_{\gamma t}(y_k) \gamma t + \mathrm{i} \tilde V_{\gamma t}(y_k)  \\
	&\qquad\qquad -\mathrm{i} \mathrm{e}^{-(\mathrm{i}y_k+1)\gamma t}  \bigg) \frac{\tilde W_{\gamma\sigma/2c}(y_k)}{y_k} \mathrm{e}^{-\gamma t} \ ,
\end{split}
\end{align}
where $\tilde W_\alpha$, $\tilde U_\alpha$, and $\tilde V_\alpha$ are given by Eqs.~(\ref{W.tilde}) and (\ref{tilde}).
The corresponding Fourier transform to real space, $\Phi_1(x,t)=(2\pi)^{-1} \int \mathrm{e}^{\mathrm{i}kx} \hat{\Phi}_1(k,t) \, \mathrm{d}k$, is given by Eq.~(\ref{damped.perturbation}).

\end{appendix}


\begin{thebibliography}{43}%
\makeatletter
\providecommand \@ifxundefined [1]{%
 \@ifx{#1\undefined}
}%
\providecommand \@ifnum [1]{%
 \ifnum #1\expandafter \@firstoftwo
 \else \expandafter \@secondoftwo
 \fi
}%
\providecommand \@ifx [1]{%
 \ifx #1\expandafter \@firstoftwo
 \else \expandafter \@secondoftwo
 \fi
}%
\providecommand \natexlab [1]{#1}%
\providecommand \enquote  [1]{``#1''}%
\providecommand \bibnamefont  [1]{#1}%
\providecommand \bibfnamefont [1]{#1}%
\providecommand \citenamefont [1]{#1}%
\providecommand \href@noop [0]{\@secondoftwo}%
\providecommand \href [0]{\begingroup \@sanitize@url \@href}%
\providecommand \@href[1]{\@@startlink{#1}\@@href}%
\providecommand \@@href[1]{\endgroup#1\@@endlink}%
\providecommand \@sanitize@url [0]{\catcode `\\12\catcode `\$12\catcode
  `\&12\catcode `\#12\catcode `\^12\catcode `\_12\catcode `\%12\relax}%
\providecommand \@@startlink[1]{}%
\providecommand \@@endlink[0]{}%
\providecommand \url  [0]{\begingroup\@sanitize@url \@url }%
\providecommand \@url [1]{\endgroup\@href {#1}{\urlprefix }}%
\providecommand \urlprefix  [0]{URL }%
\providecommand \Eprint [0]{\href }%
\providecommand \doibase [0]{http://dx.doi.org/}%
\providecommand \selectlanguage [0]{\@gobble}%
\providecommand \bibinfo  [0]{\@secondoftwo}%
\providecommand \bibfield  [0]{\@secondoftwo}%
\providecommand \translation [1]{[#1]}%
\providecommand \BibitemOpen [0]{}%
\providecommand \bibitemStop [0]{}%
\providecommand \bibitemNoStop [0]{.\EOS\space}%
\providecommand \EOS [0]{\spacefactor3000\relax}%
\providecommand \BibitemShut  [1]{\csname bibitem#1\endcsname}%
\let\auto@bib@innerbib\@empty
\bibitem [{\citenamefont {Pikovsky}, \citenamefont {Rosenblum},\ and\
  \citenamefont {Kurths}(2003)}]{Pikovsky2003}%
  \BibitemOpen
  \bibfield  {author} {\bibinfo {author} {\bibfnamefont {A.}~\bibnamefont
  {Pikovsky}}, \bibinfo {author} {\bibfnamefont {M.}~\bibnamefont {Rosenblum}},
  \ and\ \bibinfo {author} {\bibfnamefont {J.}~\bibnamefont {Kurths}},\
  }\href@noop {} {\emph {\bibinfo {title} {{Synchronization: A Universal
  Concept in Nonlinear Sciences}}}},\ {Cambridge Nonlinear Science Series}\
  (\bibinfo  {publisher} {Cambridge University Press},\ \bibinfo {year}
  {2003})\BibitemShut {NoStop}%
\bibitem [{\citenamefont {Kuramoto}(1984{\natexlab{a}})}]{Kuramoto1984}%
  \BibitemOpen
  \bibfield  {author} {\bibinfo {author} {\bibfnamefont {Y.}~\bibnamefont
  {Kuramoto}},\ }\href@noop {} {\emph {\bibinfo {title} {{Chemical
  Oscillations, Waves, and Turbulence}}}}\ (\bibinfo  {publisher}
  {Springer-Verlag},\ \bibinfo {address} {Berlin},\ \bibinfo {year}
  {1984})\BibitemShut {NoStop}%
\bibitem [{\citenamefont {Cross}\ and\ \citenamefont
  {Hohenberg}(1993)}]{Cross1993}%
  \BibitemOpen
  \bibfield  {author} {\bibinfo {author} {\bibfnamefont {M.~C.}\ \bibnamefont
  {Cross}}\ and\ \bibinfo {author} {\bibfnamefont {P.~C.}\ \bibnamefont
  {Hohenberg}},\ }\bibfield  {title} {\enquote {\bibinfo {title} {{Pattern
  formation outside of equilibrium}},}\ }\href@noop {} {\bibfield  {journal}
  {\bibinfo  {journal} {Rev. Mod. Phys.}\ }\textbf {\bibinfo {volume} {65}},\ \bibinfo {pages}
  {851}
  (\bibinfo {year} {1993})}\BibitemShut {NoStop}%
\bibitem [{\citenamefont {Strogatz}\ and\ \citenamefont
  {Mirollo}(1993)}]{Strogatz1993}%
  \BibitemOpen
  \bibfield  {author} {\bibinfo {author} {\bibfnamefont {S.~H.}\ \bibnamefont
  {Strogatz}}\ and\ \bibinfo {author} {\bibfnamefont {R.~E.}\ \bibnamefont
  {Mirollo}},\ }\bibfield  {title} {\enquote {\bibinfo {title} {{Splay states
  in globally coupled Josephson arrays: Analytical prediction of Floquet
  multipliers}},}\ }\href@noop {} {\bibfield  {journal} {\bibinfo  {journal}
  {Phys. Rev. E}\ }\textbf {\bibinfo {volume} {47}},\ \bibinfo {pages}
  {220--227} (\bibinfo {year} {1993})}\BibitemShut {NoStop}%
\bibitem [{\citenamefont {Alexeeva}, \citenamefont {Barashenkov},\ and\
  \citenamefont {Tsironis}(2000)}]{Alexeeva2000}%
  \BibitemOpen
  \bibfield  {author} {\bibinfo {author} {\bibfnamefont {N.}~\bibnamefont
  {Alexeeva}}, \bibinfo {author} {\bibfnamefont {I.}~\bibnamefont
  {Barashenkov}}, \ and\ \bibinfo {author} {\bibfnamefont {G.}~\bibnamefont
  {Tsironis}},\ }\bibfield  {title} {\enquote {\bibinfo {title}
  {{Impurity-Induced Stabilization of Solitons in Arrays of Parametrically
  Driven Nonlinear Oscillators}},}\ }\href@noop {} {\bibfield  {journal}
  {\bibinfo  {journal} {Phys. Rev. Lett.}\ }\textbf {\bibinfo {volume} {84}},\
  \bibinfo {pages} {3053--3056} (\bibinfo {year} {2000})}\BibitemShut {NoStop}%
\bibitem [{\citenamefont {Jeong}, \citenamefont {Ko},\ and\ \citenamefont
  {Moon}(2002)}]{Jeong2002}%
  \BibitemOpen
  \bibfield  {author} {\bibinfo {author} {\bibfnamefont {S.-O.}\ \bibnamefont
  {Jeong}}, \bibinfo {author} {\bibfnamefont {T.-W.}\ \bibnamefont {Ko}}, \
  and\ \bibinfo {author} {\bibfnamefont {H.-T.}\ \bibnamefont {Moon}},\
  }\bibfield  {title} {\enquote {\bibinfo {title} {{Time-Delayed Spatial
  Patterns in a Two-Dimensional Array of Coupled Oscillators}},}\ }\href@noop
  {} {\bibfield  {journal} {\bibinfo  {journal} {Phys. Rev. Lett.}\ }\textbf
  {\bibinfo {volume} {89}},\ \bibinfo {pages} {154104} (\bibinfo {year}
  {2002})}\BibitemShut {NoStop}%
\bibitem [{\citenamefont {Kim}\ \emph {et~al.}(2004)\citenamefont {Kim},
  \citenamefont {Ko}, \citenamefont {Jeong},\ and\ \citenamefont
  {Moon}}]{Kim2004}%
  \BibitemOpen
  \bibfield  {author} {\bibinfo {author} {\bibfnamefont {P.-J.}\ \bibnamefont
  {Kim}}, \bibinfo {author} {\bibfnamefont {T.-W.}\ \bibnamefont {Ko}},
  \bibinfo {author} {\bibfnamefont {H.}~\bibnamefont {Jeong}}, \ and\ \bibinfo
  {author} {\bibfnamefont {H.-T.}\ \bibnamefont {Moon}},\ }\bibfield  {title}
  {\enquote {\bibinfo {title} {{Pattern formation in a two-dimensional array of
  oscillators with phase-shifted coupling}},}\ }\href@noop {} {\bibfield
  {journal} {\bibinfo  {journal} {Phys. Rev. E}\ }\textbf {\bibinfo {volume}
  {70}},\ \bibinfo {pages} {065201} (\bibinfo {year} {2004})}\BibitemShut
  {NoStop}%
\bibitem [{\citenamefont {Ahnert}\ and\ \citenamefont
  {Pikovsky}(2008)}]{Ahnert2008}%
  \BibitemOpen
  \bibfield  {author} {\bibinfo {author} {\bibfnamefont {K.}~\bibnamefont
  {Ahnert}}\ and\ \bibinfo {author} {\bibfnamefont {A.}~\bibnamefont
  {Pikovsky}},\ }\bibfield  {title} {\enquote {\bibinfo {title} {{Traveling
  waves and compactons in phase oscillator lattices.}}}\ }\href@noop {}
  {\bibfield  {journal} {\bibinfo  {journal} {Chaos}\ }\textbf {\bibinfo
  {volume} {18}},\ \bibinfo {pages} {037118} (\bibinfo {year}
  {2008})}\BibitemShut {NoStop}%
\bibitem [{\citenamefont {Lee}\ \emph {et~al.}(2011)\citenamefont {Lee},
  \citenamefont {Restrepo}, \citenamefont {Ott},\ and\ \citenamefont
  {Antonsen}}]{Lee2011}%
  \BibitemOpen
  \bibfield  {author} {\bibinfo {author} {\bibfnamefont {W.~S.}\ \bibnamefont
  {Lee}}, \bibinfo {author} {\bibfnamefont {J.~G.}\ \bibnamefont {Restrepo}},
  \bibinfo {author} {\bibfnamefont {E.}~\bibnamefont {Ott}}, \ and\ \bibinfo
  {author} {\bibfnamefont {T.~M.}\ \bibnamefont {Antonsen}},\ }\bibfield
  {title} {\enquote {\bibinfo {title} {{Dynamics and pattern formation in large
  systems of spatially-coupled oscillators with finite response times.}}}\
  }\href@noop {} {\bibfield  {journal} {\bibinfo  {journal} {Chaos}\ }\textbf
  {\bibinfo {volume} {21}},\ \bibinfo {pages} {023122} (\bibinfo {year}
  {2011})}\BibitemShut {NoStop}%
\bibitem [{\citenamefont {Ares}\ \emph {et~al.}(2012)\citenamefont {Ares},
  \citenamefont {Morelli}, \citenamefont {J\"org}, \citenamefont {Oates},\ and\
  \citenamefont {J\"ulicher}}]{Ares2012}%
  \BibitemOpen
  \bibfield  {author} {\bibinfo {author} {\bibfnamefont {S.}~\bibnamefont
  {Ares}}, \bibinfo {author} {\bibfnamefont {L.~G.}\ \bibnamefont {Morelli}},
  \bibinfo {author} {\bibfnamefont {D.~J.}\ \bibnamefont {J\"org}}, \bibinfo
  {author} {\bibfnamefont {A.~C.}\ \bibnamefont {Oates}}, \ and\ \bibinfo
  {author} {\bibfnamefont {F.}~\bibnamefont {J\"ulicher}},\ }\bibfield  {title}
  {\enquote {\bibinfo {title} {{Collective Modes of Coupled Phase Oscillators
  with Delayed Coupling}},}\ }\href@noop {} {\bibfield  {journal} {\bibinfo
  {journal} {Phys. Rev. Lett.}\ }\textbf {\bibinfo {volume} {108}},\ \bibinfo
  {pages} {204101} (\bibinfo {year} {2012})}\BibitemShut {NoStop}%
\bibitem [{\citenamefont {J\"{o}rg}\ \emph {et~al.}(2014)\citenamefont
  {J\"{o}rg}, \citenamefont {Morelli}, \citenamefont {Ares},\ and\
  \citenamefont {J\"{u}licher}}]{Jorg2014}%
  \BibitemOpen
  \bibfield  {author} {\bibinfo {author} {\bibfnamefont {D.~J.}\ \bibnamefont
  {J\"{o}rg}}, \bibinfo {author} {\bibfnamefont {L.~G.}\ \bibnamefont
  {Morelli}}, \bibinfo {author} {\bibfnamefont {S.}~\bibnamefont {Ares}}, \
  and\ \bibinfo {author} {\bibfnamefont {F.}~\bibnamefont {J\"{u}licher}},\
  }\bibfield  {title} {\enquote {\bibinfo {title} {{Synchronization in the
  Presence of Phase Shifts and Coupling Delays}},}\ }\href@noop {} {\bibfield
  {journal} {\bibinfo  {journal} {Phys. Rev. Lett.}\ }\textbf {\bibinfo
  {volume} {112}},\ \bibinfo {pages} {174101} (\bibinfo {year}
  {2014})}\BibitemShut {NoStop}%
\bibitem [{\citenamefont {Ortoleva}\ and\ \citenamefont
  {Ross}(1974)}]{Ortoleva1974}%
  \BibitemOpen
  \bibfield  {author} {\bibinfo {author} {\bibfnamefont {P.}~\bibnamefont
  {Ortoleva}}\ and\ \bibinfo {author} {\bibfnamefont {J.}~\bibnamefont
  {Ross}},\ }\bibfield  {title} {\enquote {\bibinfo {title} {{On a variety of
  wave phenomena in chemical reactions}},}\ }\href@noop {} {\bibfield
  {journal} {\bibinfo  {journal} {J. Chem. Phys.}\ }\textbf {\bibinfo {volume}
  {60}},\ \bibinfo {pages} {5090} (\bibinfo {year} {1974})}\BibitemShut
  {NoStop}%
\bibitem [{\citenamefont {Sakaguchi}, \citenamefont {Shinomoto},\ and\
  \citenamefont {Kuramoto}(1988)}]{Sakaguchi1988}%
  \BibitemOpen
  \bibfield  {author} {\bibinfo {author} {\bibfnamefont {H.}~\bibnamefont
  {Sakaguchi}}, \bibinfo {author} {\bibfnamefont {S.}~\bibnamefont
  {Shinomoto}}, \ and\ \bibinfo {author} {\bibfnamefont {Y.}~\bibnamefont
  {Kuramoto}},\ }\bibfield  {title} {\enquote {\bibinfo {title} {{Mutual
  Entrainment in Oscillator Lattices with Nonvariational Type Interaction}},}\
  }\href@noop {} {\bibfield  {journal} {\bibinfo  {journal} {Prog. Theor.
  Phys.}\ }\textbf {\bibinfo {volume} {79}},\ \bibinfo {pages} {1069--1079}
  (\bibinfo {year} {1988})}\BibitemShut {NoStop}%
\bibitem [{\citenamefont {Rosenau}\ and\ \citenamefont
  {Pikovsky}(2005)}]{Rosenau2005}%
  \BibitemOpen
  \bibfield  {author} {\bibinfo {author} {\bibfnamefont {P.}~\bibnamefont
  {Rosenau}}\ and\ \bibinfo {author} {\bibfnamefont {A.}~\bibnamefont
  {Pikovsky}},\ }\bibfield  {title} {\enquote {\bibinfo {title} {{Phase
  Compactons in Chains of Dispersively Coupled Oscillators}},}\ }\href@noop {}
  {\bibfield  {journal} {\bibinfo  {journal} {Phys. Rev. Lett.}\ }\textbf
  {\bibinfo {volume} {94}},\ \bibinfo {pages} {174102} (\bibinfo {year}
  {2005})}\BibitemShut {NoStop}%
\bibitem [{\citenamefont {Murray}, \citenamefont {Maini},\ and\ \citenamefont
  {Baker}(2011)}]{Murray2011}%
  \BibitemOpen
  \bibfield  {author} {\bibinfo {author} {\bibfnamefont {P.~J.}\ \bibnamefont
  {Murray}}, \bibinfo {author} {\bibfnamefont {P.~K.}\ \bibnamefont {Maini}}, \
  and\ \bibinfo {author} {\bibfnamefont {R.~E.}\ \bibnamefont {Baker}},\
  }\bibfield  {title} {\enquote {\bibinfo {title} {{The clock and wavefront
  model revisited}},}\ }\href@noop {} {\bibfield  {journal} {\bibinfo
  {journal} {J. Theor. Biol.}\ }\textbf {\bibinfo {volume} {283}},\ \bibinfo
  {pages} {227--38} (\bibinfo {year} {2011})}\BibitemShut {NoStop}%
\bibitem [{\citenamefont {Hoang}, \citenamefont {Jo},\ and\ \citenamefont
  {Hong}(2015)}]{Hoang2015}%
  \BibitemOpen
  \bibfield  {author} {\bibinfo {author} {\bibfnamefont {D.-T.}\ \bibnamefont
  {Hoang}}, \bibinfo {author} {\bibfnamefont {J.}~\bibnamefont {Jo}}, \ and\
  \bibinfo {author} {\bibfnamefont {H.}~\bibnamefont {Hong}},\ }\bibfield
  {title} {\enquote {\bibinfo {title} {{Traveling wave in a three-dimensional
  array of conformist and contrarian oscillators}},}\ }\href@noop {} {\bibfield
   {journal} {\bibinfo  {journal} {Phys. Rev. E}\ }\textbf {\bibinfo {volume}
  {91}},\ \bibinfo {pages} {032135} (\bibinfo {year} {2015})}\BibitemShut
  {NoStop}%
\bibitem [{\citenamefont {Morelli}\ \emph {et~al.}(2009)\citenamefont
  {Morelli}, \citenamefont {Ares}, \citenamefont {Herrgen}, \citenamefont
  {Schr{\"o}ter}, \citenamefont {J{\"u}licher},\ and\ \citenamefont
  {Oates}}]{Morelli2009}%
  \BibitemOpen
  \bibfield  {author} {\bibinfo {author} {\bibfnamefont {L.~G.}\ \bibnamefont
  {Morelli}}, \bibinfo {author} {\bibfnamefont {S.}~\bibnamefont {Ares}},
  \bibinfo {author} {\bibfnamefont {L.}~\bibnamefont {Herrgen}}, \bibinfo
  {author} {\bibfnamefont {C.}~\bibnamefont {Schr{\"o}ter}}, \bibinfo {author}
  {\bibfnamefont {F.}~\bibnamefont {J{\"u}licher}}, \ and\ \bibinfo {author}
  {\bibfnamefont {A.~C.}\ \bibnamefont {Oates}},\ }\bibfield  {title} {\enquote
  {\bibinfo {title} {{Delayed coupling theory of vertebrate segmentation}},}\
  }\href@noop {} {\bibfield  {journal} {\bibinfo  {journal} {HFSP J.}\ }\textbf
  {\bibinfo {volume} {3}},\ \bibinfo {pages} {55} (\bibinfo {year}
  {2009})}\BibitemShut {NoStop}%
\bibitem [{\citenamefont {Szwaj}, \citenamefont {Bielawski},\ and\
  \citenamefont {Derozier}(1996)}]{Szwaj1996}%
  \BibitemOpen
  \bibfield  {author} {\bibinfo {author} {\bibfnamefont {C.}~\bibnamefont
  {Szwaj}}, \bibinfo {author} {\bibfnamefont {S.}~\bibnamefont {Bielawski}}, \
  and\ \bibinfo {author} {\bibfnamefont {D.}~\bibnamefont {Derozier}},\
  }\bibfield  {title} {\enquote {\bibinfo {title} {{Propagation of Waves in the
  Spectrum of a Multimode Laser}},}\ }\href@noop {} {\bibfield  {journal}
  {\bibinfo  {journal} {Phys. Rev. Lett.}\ }\textbf {\bibinfo {volume} {77}},\
  \bibinfo {pages} {4540--4543} (\bibinfo {year} {1996})}\BibitemShut {NoStop}%
\bibitem [{\citenamefont {Mat\'{\i}as}\ and\ \citenamefont
  {G\"{u}\'{e}mez}(1998)}]{Matias1998}%
  \BibitemOpen
  \bibfield  {author} {\bibinfo {author} {\bibfnamefont {M.}~\bibnamefont
  {Mat\'{\i}as}}\ and\ \bibinfo {author} {\bibfnamefont {J.}~\bibnamefont
  {G\"{u}\'{e}mez}},\ }\bibfield  {title} {\enquote {\bibinfo {title}
  {{Transient Periodic Rotating Waves and Fast Propagation of Synchronization
  in Linear Arrays of Chaotic Systems}},}\ }\href@noop {} {\bibfield  {journal}
  {\bibinfo  {journal} {Phys. Rev. Lett.}\ }\textbf {\bibinfo {volume} {81}},\
  \bibinfo {pages} {4124--4127} (\bibinfo {year} {1998})}\BibitemShut {NoStop}%
\bibitem [{\citenamefont {Horikawa}\ and\ \citenamefont
  {Kitajima}(2012)}]{Horikawa2012}%
  \BibitemOpen
  \bibfield  {author} {\bibinfo {author} {\bibfnamefont {Y.}~\bibnamefont
  {Horikawa}}\ and\ \bibinfo {author} {\bibfnamefont {H.}~\bibnamefont
  {Kitajima}},\ }\bibfield  {title} {\enquote {\bibinfo {title} {{Quasiperiodic
  and exponential transient phase waves and their bifurcations in a ring of
  unidirectionally coupled parametric oscillators}},}\ }\href@noop {}
  {\bibfield  {journal} {\bibinfo  {journal} {Nonlin. Dyn.}\ }\textbf {\bibinfo
  {volume} {70}},\ \bibinfo {pages} {1079--1094} (\bibinfo {year}
  {2012})}\BibitemShut {NoStop}%
\bibitem [{\citenamefont {Tanaka}, \citenamefont {Lichtenberg},\ and\
  \citenamefont {Oishi}(1997{\natexlab{a}})}]{Tanaka1997a}%
  \BibitemOpen
  \bibfield  {author} {\bibinfo {author} {\bibfnamefont {H.-A.}\ \bibnamefont
  {Tanaka}}, \bibinfo {author} {\bibfnamefont {A.}~\bibnamefont {Lichtenberg}},
  \ and\ \bibinfo {author} {\bibfnamefont {S.}~\bibnamefont {Oishi}},\
  }\bibfield  {title} {\enquote {\bibinfo {title} {{First Order Phase
  Transition Resulting from Finite Inertia in Coupled Oscillator Systems}},}\
  }\href@noop {} {\bibfield  {journal} {\bibinfo  {journal} {Phys. Rev. Lett.}\
  }\textbf {\bibinfo {volume} {78}},\ \bibinfo {pages} {2104--2107} (\bibinfo
  {year} {1997}{\natexlab{a}})}\BibitemShut {NoStop}%
\bibitem [{\citenamefont {Acebr\'{o}n}\ and\ \citenamefont
  {Spigler}(1998)}]{Acebron1998}%
  \BibitemOpen
  \bibfield  {author} {\bibinfo {author} {\bibfnamefont {J.~A.}\ \bibnamefont
  {Acebr\'{o}n}}\ and\ \bibinfo {author} {\bibfnamefont {R.}~\bibnamefont
  {Spigler}},\ }\bibfield  {title} {\enquote {\bibinfo {title} {{Adaptive
  frequency model for phase-frequency synchronization in large populations of
  globally coupled nonlinear oscillators}},}\ }\href@noop {} {\bibfield
  {journal} {\bibinfo  {journal} {Phys. Rev. Lett.}\ }\textbf {\bibinfo
  {volume} {81}},\ \bibinfo {pages} {2229--2232} (\bibinfo {year}
  {1998})}\BibitemShut {NoStop}%
\bibitem [{\citenamefont {Choi}, \citenamefont {Ha},\ and\ \citenamefont
  {Noh}(2013)}]{Choi2013}%
  \BibitemOpen
  \bibfield  {author} {\bibinfo {author} {\bibfnamefont {Y.-P.}\ \bibnamefont
  {Choi}}, \bibinfo {author} {\bibfnamefont {S.-Y.}\ \bibnamefont {Ha}}, \ and\
  \bibinfo {author} {\bibfnamefont {S.~E.}\ \bibnamefont {Noh}},\ }\bibfield
  {title} {\enquote {\bibinfo {title} {{On the relaxation dynamics of the
  Kuramoto oscillators with small inertia}},}\ }\href@noop {} {\bibfield
  {journal} {\bibinfo  {journal} {J. Math. Phys.}\ }\textbf {\bibinfo {volume}
  {54}},\ \bibinfo {pages} {072701} (\bibinfo {year} {2013})}\BibitemShut
  {NoStop}%
\bibitem [{\citenamefont {Olmi}\ \emph {et~al.}(2014)\citenamefont {Olmi},
  \citenamefont {Navas}, \citenamefont {Boccaletti},\ and\ \citenamefont
  {Torcini}}]{Olmi2014}%
  \BibitemOpen
  \bibfield  {author} {\bibinfo {author} {\bibfnamefont {S.}~\bibnamefont
  {Olmi}}, \bibinfo {author} {\bibfnamefont {A.}~\bibnamefont {Navas}},
  \bibinfo {author} {\bibfnamefont {S.}~\bibnamefont {Boccaletti}}, \ and\
  \bibinfo {author} {\bibfnamefont {A.}~\bibnamefont {Torcini}},\ }\bibfield
  {title} {\enquote {\bibinfo {title} {{Hysteretic transitions in the Kuramoto
  model with inertia}},}\ }\href@noop {} {\bibfield  {journal} {\bibinfo
  {journal} {Phys. Rev. E}\ }\textbf {\bibinfo {volume} {90}},\ \bibinfo
  {pages} {042905} (\bibinfo {year} {2014})}\BibitemShut {NoStop}%
\bibitem [{\citenamefont {Ermentrout}(1991)}]{Ermentrout1991}%
  \BibitemOpen
  \bibfield  {author} {\bibinfo {author} {\bibfnamefont {B.}~\bibnamefont
  {Ermentrout}},\ }\bibfield  {title} {\enquote {\bibinfo {title} {{An adaptive
  model for synchrony in the firefly pteroptyx malaccae}},}\ }\href@noop {}
  {\bibfield  {journal} {\bibinfo  {journal} {J. Math. Biol.}\ }\textbf
  {\bibinfo {volume} {29}},\ \bibinfo {pages} {571--585} (\bibinfo {year}
  {1991})}\BibitemShut {NoStop}%
\bibitem [{\citenamefont {Tanaka}, \citenamefont {Lichtenberg},\ and\
  \citenamefont {Oishi}(1997{\natexlab{b}})}]{Tanaka1997b}%
  \BibitemOpen
  \bibfield  {author} {\bibinfo {author} {\bibfnamefont {H.-A.}\ \bibnamefont
  {Tanaka}}, \bibinfo {author} {\bibfnamefont {A.~J.}\ \bibnamefont
  {Lichtenberg}}, \ and\ \bibinfo {author} {\bibfnamefont {S.}~\bibnamefont
  {Oishi}},\ }\bibfield  {title} {\enquote {\bibinfo {title}
  {{Self-synchronization of coupled oscillators with hysteretic responses}},}\
  }\href@noop {} {\bibfield  {journal} {\bibinfo  {journal} {Physica D}\
  }\textbf {\bibinfo {volume} {100}},\ \bibinfo {pages} {279--300} (\bibinfo
  {year} {1997}{\natexlab{b}})}\BibitemShut {NoStop}%
\bibitem [{\citenamefont {Hong}, \citenamefont {Jeon},\ and\ \citenamefont
  {Choi}(2002)}]{Hong2002}%
  \BibitemOpen
  \bibfield  {author} {\bibinfo {author} {\bibfnamefont {H.}~\bibnamefont
  {Hong}}, \bibinfo {author} {\bibfnamefont {G.}~\bibnamefont {Jeon}}, \ and\
  \bibinfo {author} {\bibfnamefont {M.}~\bibnamefont {Choi}},\ }\bibfield
  {title} {\enquote {\bibinfo {title} {{Spontaneous phase oscillation induced
  by inertia and time delay}},}\ }\href@noop {} {\bibfield  {journal} {\bibinfo
   {journal} {Phys. Rev. E}\ }\textbf {\bibinfo {volume} {65}},\ \bibinfo
  {pages} {026208} (\bibinfo {year} {2002})}\BibitemShut {NoStop}%
\bibitem [{\citenamefont {Dolan}, \citenamefont {Majtanik},\ and\ \citenamefont
  {Tass}(2005)}]{Dolan2005}%
  \BibitemOpen
  \bibfield  {author} {\bibinfo {author} {\bibfnamefont {K.}~\bibnamefont
  {Dolan}}, \bibinfo {author} {\bibfnamefont {M.}~\bibnamefont {Majtanik}}, \
  and\ \bibinfo {author} {\bibfnamefont {P.~A.}\ \bibnamefont {Tass}},\
  }\bibfield  {title} {\enquote {\bibinfo {title} {{Phase resetting and
  transient desynchronization in networks of globally coupled phase oscillators
  with inertia}},}\ }\href@noop {} {\bibfield  {journal} {\bibinfo  {journal}
  {Physica D}\ }\textbf {\bibinfo {volume} {211}},\ \bibinfo {pages} {128--138}
  (\bibinfo {year} {2005})}\BibitemShut {NoStop}%
\bibitem [{\citenamefont {Ji}\ \emph {et~al.}(2013)\citenamefont {Ji},
  \citenamefont {Peron}, \citenamefont {Menck}, \citenamefont {Rodrigues},\
  and\ \citenamefont {Kurths}}]{Ji2013}%
  \BibitemOpen
  \bibfield  {author} {\bibinfo {author} {\bibfnamefont {P.}~\bibnamefont
  {Ji}}, \bibinfo {author} {\bibfnamefont {T.~K.~D.}\ \bibnamefont {Peron}},
  \bibinfo {author} {\bibfnamefont {P.~J.}\ \bibnamefont {Menck}}, \bibinfo
  {author} {\bibfnamefont {F.~A.}\ \bibnamefont {Rodrigues}}, \ and\ \bibinfo
  {author} {\bibfnamefont {J.}~\bibnamefont {Kurths}},\ }\bibfield  {title}
  {\enquote {\bibinfo {title} {{Cluster Explosive Synchronization in Complex
  Networks}},}\ }\href@noop {} {\bibfield  {journal} {\bibinfo  {journal}
  {Phys. Rev. Lett.}\ }\textbf {\bibinfo {volume} {110}},\ \bibinfo {pages}
  {218701} (\bibinfo {year} {2013})}\BibitemShut {NoStop}%
\bibitem [{\citenamefont {Gupta}, \citenamefont {Campa},\ and\ \citenamefont
  {Ruffo}(2014)}]{Gupta2014}%
  \BibitemOpen
  \bibfield  {author} {\bibinfo {author} {\bibfnamefont {S.}~\bibnamefont
  {Gupta}}, \bibinfo {author} {\bibfnamefont {A.}~\bibnamefont {Campa}}, \ and\
  \bibinfo {author} {\bibfnamefont {S.}~\bibnamefont {Ruffo}},\ }\bibfield
  {title} {\enquote {\bibinfo {title} {{Nonequilibrium first-order phase
  transition in coupled oscillator systems with inertia and noise}},}\
  }\href@noop {} {\bibfield  {journal} {\bibinfo  {journal} {Phys. Rev. E}\
  }\textbf {\bibinfo {volume} {89}},\ \bibinfo {pages} {022123} (\bibinfo
  {year} {2014})}\BibitemShut {NoStop}%
\bibitem [{\citenamefont {Komarov}, \citenamefont {Gupta},\ and\ \citenamefont
  {Pikovsky}(2014)}]{Komarov2014}%
  \BibitemOpen
  \bibfield  {author} {\bibinfo {author} {\bibfnamefont {M.}~\bibnamefont
  {Komarov}}, \bibinfo {author} {\bibfnamefont {S.}~\bibnamefont {Gupta}}, \
  and\ \bibinfo {author} {\bibfnamefont {A.}~\bibnamefont {Pikovsky}},\
  }\bibfield  {title} {\enquote {\bibinfo {title} {{Synchronization transitions
  in globally coupled rotors in the presence of noise and inertia: Exact
  results}},}\ }\href@noop {} {\bibfield  {journal} {\bibinfo  {journal} {EPL
  (Europhysics Letters)}\ }\textbf {\bibinfo {volume} {106}},\ \bibinfo {pages}
  {40003} (\bibinfo {year} {2014})}\BibitemShut {NoStop}%
\bibitem [{\citenamefont {Kuramoto}(1984{\natexlab{b}})}]{Kuramoto1984b}%
  \BibitemOpen
  \bibfield  {author} {\bibinfo {author} {\bibfnamefont {Y.}~\bibnamefont
  {Kuramoto}},\ }\bibfield  {title} {\enquote {\bibinfo {title} {{Cooperative
  Dynamics of Oscillator Community}},}\ }\href@noop {} {\bibfield  {journal}
  {\bibinfo  {journal} {Prog. Theor. Phys.}\ }\textbf {\bibinfo {volume}
  {79}},\ \bibinfo {pages} {223--240} (\bibinfo {year}
  {1984}{\natexlab{b}})}\BibitemShut {NoStop}%
\bibitem [{\citenamefont {Zauderer}(2006)}]{Zauderer2006}%
  \BibitemOpen
  \bibfield  {author} {\bibinfo {author} {\bibfnamefont {E.}~\bibnamefont
  {Zauderer}},\ }\href@noop {} {\emph {\bibinfo {title} {{Partial Differential
  Equations of Applied Mathematics}}}}\ (\bibinfo  {publisher} {John Wiley \&
  Sons},\ \bibinfo {year} {2006})\BibitemShut {NoStop}%
\bibitem [{\citenamefont {Choe}\ \emph {et~al.}(2010)\citenamefont {Choe},
  \citenamefont {Dahms}, \citenamefont {H\"{o}vel},\ and\ \citenamefont
  {Sch\"{o}ll}}]{Choe2010}%
  \BibitemOpen
  \bibfield  {author} {\bibinfo {author} {\bibfnamefont {C.-U.}\ \bibnamefont
  {Choe}}, \bibinfo {author} {\bibfnamefont {T.}~\bibnamefont {Dahms}},
  \bibinfo {author} {\bibfnamefont {P.}~\bibnamefont {H\"{o}vel}}, \ and\
  \bibinfo {author} {\bibfnamefont {E.}~\bibnamefont {Sch\"{o}ll}},\ }\bibfield
   {title} {\enquote {\bibinfo {title} {{Controlling synchrony by delay
  coupling in networks: From in-phase to splay and cluster states}},}\
  }\href@noop {} {\bibfield  {journal} {\bibinfo  {journal} {Phys. Rev. E}\
  }\textbf {\bibinfo {volume} {81}},\ \bibinfo {pages} {025205} (\bibinfo
  {year} {2010})}\BibitemShut {NoStop}%
\bibitem [{\citenamefont {Bender}\ and\ \citenamefont
  {Orszag}(1999)}]{Bender1999}%
  \BibitemOpen
  \bibfield  {author} {\bibinfo {author} {\bibfnamefont {C.~M.}\ \bibnamefont
  {Bender}}\ and\ \bibinfo {author} {\bibfnamefont {S.~A.}\ \bibnamefont
  {Orszag}},\ }\href@noop {} {\emph {\bibinfo {title} {{Advanced Mathematical
  Methods for Scientists and Engineers}}}}\ (\bibinfo  {publisher} {Springer},\
  \bibinfo {year} {1999})\BibitemShut {NoStop}%
\bibitem [{\citenamefont {Kardar}, \citenamefont {Parisi},\ and\ \citenamefont
  {Zhang}(1986)}]{Kardar1986}%
  \BibitemOpen
  \bibfield  {author} {\bibinfo {author} {\bibfnamefont {M.}~\bibnamefont
  {Kardar}}, \bibinfo {author} {\bibfnamefont {G.}~\bibnamefont {Parisi}}, \
  and\ \bibinfo {author} {\bibfnamefont {Y.~C.}\ \bibnamefont {Zhang}},\
  }\bibfield  {title} {\enquote {\bibinfo {title} {{Dynamic Scaling of Growing
  Interfaces}},}\ }\href@noop {} {\bibfield  {journal} {\bibinfo  {journal}
  {Phys. Rev. Lett.}\ }\textbf {\bibinfo {volume} {56}},\ \bibinfo {pages}
  {889} (\bibinfo {year} {1986})}\BibitemShut {NoStop}%
\bibitem [{\citenamefont {Kar}, \citenamefont {Banik},\ and\ \citenamefont
  {Ray}(2003)}]{Kar2003}%
  \BibitemOpen
  \bibfield  {author} {\bibinfo {author} {\bibfnamefont {S.}~\bibnamefont
  {Kar}}, \bibinfo {author} {\bibfnamefont {S.~K.}\ \bibnamefont {Banik}}, \
  and\ \bibinfo {author} {\bibfnamefont {D.~S.}\ \bibnamefont {Ray}},\
  }\bibfield  {title} {\enquote {\bibinfo {title} {{Exact solutions of Fisher
  and Burgers equations with finite transport memory}},}\ }\href@noop {}
  {\bibfield  {journal} {\bibinfo  {journal} {J. Phys. A}\ }\textbf {\bibinfo
  {volume} {36}},\ \bibinfo {pages} {2771--2780} (\bibinfo {year}
  {2003})}\BibitemShut {NoStop}%
\bibitem [{\citenamefont {Thorp}, \citenamefont {Seyler},\ and\ \citenamefont
  {Phadke}(1998)}]{Thorp1998}%
  \BibitemOpen
  \bibfield  {author} {\bibinfo {author} {\bibfnamefont {J.~S.}\ \bibnamefont
  {Thorp}}, \bibinfo {author} {\bibfnamefont {C.~E.}\ \bibnamefont {Seyler}}, \
  and\ \bibinfo {author} {\bibfnamefont {A.~G.}\ \bibnamefont {Phadke}},\
  }\bibfield  {title} {\enquote {\bibinfo {title} {{Electromechanical wave
  propagation in large electric power systems}},}\ }\href@noop {} {\bibfield
  {journal} {\bibinfo  {journal} {IEEE Transactions on Circuits and Systems I:
  Fundamental Theory and Applications}\ }\textbf {\bibinfo {volume} {45}},\
  \bibinfo {pages} {614--622} (\bibinfo {year} {1998})}\BibitemShut {NoStop}%
\bibitem [{\citenamefont {Xu}\ \emph {et~al.}(2014)\citenamefont {Xu},
  \citenamefont {Wen}, \citenamefont {Ledwich},\ and\ \citenamefont
  {Xue}}]{Xu2014}%
  \BibitemOpen
  \bibfield  {author} {\bibinfo {author} {\bibfnamefont {Y.}~\bibnamefont
  {Xu}}, \bibinfo {author} {\bibfnamefont {F.}~\bibnamefont {Wen}}, \bibinfo
  {author} {\bibfnamefont {G.}~\bibnamefont {Ledwich}}, \ and\ \bibinfo
  {author} {\bibfnamefont {Y.}~\bibnamefont {Xue}},\ }\bibfield  {title}
  {\enquote {\bibinfo {title} {{Electromechanical wave in power systems: theory
  and applications}},}\ }\href@noop {} {\bibfield  {journal} {\bibinfo
  {journal} {J. Mod. Power Syst. Clean Energy}\ }\textbf {\bibinfo {volume}
  {2}},\ \bibinfo {pages} {163--172} (\bibinfo {year} {2014})}\BibitemShut
  {NoStop}%
\bibitem [{\citenamefont {Filatrella}, \citenamefont {Nielsen},\ and\
  \citenamefont {Pedersen}(2008)}]{Filatrella2008}%
  \BibitemOpen
  \bibfield  {author} {\bibinfo {author} {\bibfnamefont {G.}~\bibnamefont
  {Filatrella}}, \bibinfo {author} {\bibfnamefont {A.~H.}\ \bibnamefont
  {Nielsen}}, \ and\ \bibinfo {author} {\bibfnamefont {N.~F.}\ \bibnamefont
  {Pedersen}},\ }\bibfield  {title} {\enquote {\bibinfo {title} {{Analysis of a
  power grid using a Kuramoto-like model}},}\ }\href@noop {} {\bibfield
  {journal} {\bibinfo  {journal} {Eur. Phys. J. B.}\ }\textbf {\bibinfo
  {volume} {61}},\ \bibinfo {pages} {485--491} (\bibinfo {year}
  {2008})}\BibitemShut {NoStop}%
\bibitem [{\citenamefont {Rohden}\ \emph {et~al.}(2012)\citenamefont {Rohden},
  \citenamefont {Sorge}, \citenamefont {Timme},\ and\ \citenamefont
  {Witthaut}}]{Rohden2012}%
  \BibitemOpen
  \bibfield  {author} {\bibinfo {author} {\bibfnamefont {M.}~\bibnamefont
  {Rohden}}, \bibinfo {author} {\bibfnamefont {A.}~\bibnamefont {Sorge}},
  \bibinfo {author} {\bibfnamefont {M.}~\bibnamefont {Timme}}, \ and\ \bibinfo
  {author} {\bibfnamefont {D.}~\bibnamefont {Witthaut}},\ }\bibfield  {title}
  {\enquote {\bibinfo {title} {{Self-Organized Synchronization in Decentralized
  Power Grids}},}\ }\href@noop {} {\bibfield  {journal} {\bibinfo  {journal}
  {Phys. Rev. Lett.}\ }\textbf {\bibinfo {volume} {109}},\ \bibinfo {pages}
  {064101} (\bibinfo {year} {2012})}\BibitemShut {NoStop}%
\bibitem [{\citenamefont {Pollakis}\ \emph {et~al.}(2014)\citenamefont
  {Pollakis}, \citenamefont {Wetzel}, \citenamefont {J\"{o}rg}, \citenamefont
  {Rave}, \citenamefont {Fettweis},\ and\ \citenamefont
  {J\"{u}licher}}]{Pollakis2014}%
  \BibitemOpen
  \bibfield  {author} {\bibinfo {author} {\bibfnamefont {A.}~\bibnamefont
  {Pollakis}}, \bibinfo {author} {\bibfnamefont {L.}~\bibnamefont {Wetzel}},
  \bibinfo {author} {\bibfnamefont {D.~J.}\ \bibnamefont {J\"{o}rg}}, \bibinfo
  {author} {\bibfnamefont {W.}~\bibnamefont {Rave}}, \bibinfo {author}
  {\bibfnamefont {G.}~\bibnamefont {Fettweis}}, \ and\ \bibinfo {author}
  {\bibfnamefont {F.}~\bibnamefont {J\"{u}licher}},\ }\bibfield  {title}
  {\enquote {\bibinfo {title} {{Synchronization in networks of mutually
  delay-coupled phase-locked loops}},}\ }\href@noop {} {\bibfield  {journal}
  {\bibinfo  {journal} {New J. Phys.}\ }\textbf {\bibinfo {volume} {16}},\
  \bibinfo {pages} {113009} (\bibinfo {year} {2014})}\BibitemShut {NoStop}%
\bibitem [{\citenamefont {Earl}\ and\ \citenamefont
  {Strogatz}(2003)}]{EarlStrogatz2003}%
  \BibitemOpen
  \bibfield  {author} {\bibinfo {author} {\bibfnamefont {M.~G.}\ \bibnamefont
  {Earl}}\ and\ \bibinfo {author} {\bibfnamefont {S.~H.}\ \bibnamefont
  {Strogatz}},\ }\bibfield  {title} {\enquote {\bibinfo {title}
  {{Synchronization in oscillator networks with delayed coupling: A stability
  criterion}},}\ }\href@noop {} {\bibfield  {journal} {\bibinfo  {journal}
  {Phys. Rev. E}\ }\textbf {\bibinfo {volume} {67}},\ \bibinfo {pages} {036204}
  (\bibinfo {year} {2003})}\BibitemShut {NoStop}%
\end{thebibliography}
\end{document}